\documentclass[preprintnumbers,nofootinbib,noshowpacs,eqsecnum,prd,superscriptaddress,letterpaper]{revtex4-2}
\pdfoutput=1
\usepackage{graphicx}
\usepackage{amsmath,amssymb}
\usepackage{url}
\usepackage{multirow}
\usepackage{hyperref,color}
\usepackage[normalem]{ulem}
\usepackage{slashed}
\usepackage{rotating}
\usepackage{afterpage}
\usepackage{ifthen}
\usepackage{mciteplus}
\usepackage{multirow}
\usepackage{color}
\usepackage{pifont}

\newcommand {\cyan} {\color{cyan}}

\def\lsim{\raise0.3ex\hbox{$\;<$\kern-0.75em\raise-1.1ex\hbox{$\sim\;$}}}
\def\gsim{\raise0.3ex\hbox{$\;>$\kern-0.75em\raise-1.1ex\hbox{$\sim\;$}}}

\newcommand{\Shat}{\widehat{S}}
\newcommand{\Uhat}{\widehat{U}}
\newcommand{\That}{\widehat{T}}
\newcommand{\dprime}{\prime\prime}
\newcommand{\tprime}{\prime\prime\prime}
\newcommand{\PropDer}[2][1]{\Pi_{#1}^{#2}(0)}

\newcommand{\eh}{\hat{e}}
\newcommand{\sh}{\hat{s}}
\newcommand{\ch}{\hat{c}}
\newcommand{\vh}{\hat{v}}

\newcommand {\qbw}     {{Q}_{BW}}
\newcommand {\qpone}   {{Q}_{\Phi,1}}

\newcommand {\qww}     {{Q}_{WW}}
\newcommand {\qbb}     {{Q}_{BB}}

\newcommand{\chm}{\checkmark }
\newcommand{\xm}{\ding{53}}

\begin{document}

\preprint{YITP-SB-2025-07, UWThPh 2025-12}
\title{Drell-Yan production in universal theories beyond dimension-six SMEFT}
\author{Tyler Corbett}
\email{corbett.t.s@gmail.com}
\affiliation{Faculty of Physics, University of Vienna, Boltzmanngasse 5, A-1090 Wien, Austria}
\author{Jay Desai}
\email{jay.desai@stonybrook.edu}
\affiliation{C.N. Yang Institute for Theoretical Physics,
  Stony Brook University, Stony Brook New York 11794-3849, USA}
\author{O.\ J.\ P.\ \'Eboli}
\email{eboli@if.usp.br}
\affiliation{Instituto de F\'{\i}sica, 
Universidade de S\~ao Paulo, S\~ao Paulo -- SP  05508-090, Brazil.}

\author{M.~C.~Gonzalez-Garcia}
\email{concha.gonzalez-garcia@stonybrook.edu}
\affiliation{C.N. Yang Institute for Theoretical Physics,
  Stony Brook University, Stony Brook New York 11794-3849, USA}
\affiliation{Departament de Fis\'{\i}ca Qu\`antica i
  Astrof\'{\i}sica and Institut de Ciencies del Cosmos, Universitat de
  Barcelona, Diagonal 647, E-08028 Barcelona, Spain}
\affiliation{Instituci\'o Catalana de Recerca i Estudis
  Avan\c{c}ats (ICREA), Pg. Lluis Companys 23, 08010 Barcelona,
  Spain.}
\author{Matheus Martines}
\email{matheus.martines.silva@usp.br}
\affiliation{Instituto de F\'{\i}sica, 
Universidade de S\~ao Paulo, S\~ao Paulo -- SP  05508-090, Brazil.}
\affiliation{Universit\'e Paris-Saclay, CNRS/IN2P3, IJCLab, 91405 Orsay, France}
\author{Peter Reimitz}
\email{peter@if.usp.br}
\affiliation{Instituto de F\'{\i}sica, 
Universidade de S\~ao Paulo, S\~ao Paulo -- SP  05508-090, Brazil.}
%---affiliations
\begin{abstract}
  We study Drell-Yan production in universal theories consistently
  including effects beyond dimension six in the SMEFT. Within
  universal SMEFT and with $C$ and $P$ conservation we find that
  eleven dimension-eight operators contribute in addition to the six
  contributing at dimension-six. We first work in an operator basis in
  which operators with higher derivatives of the bosonic fields have
  been rotated by equations of motion in favor of combinations of
  operators involving SM fermion currents. We derive the general form
  of the amplitudes consistently in the expansion to
  ${\cal O}(\Lambda^{-4})$ and identify eight combinations of the 17
  Wilson coefficients which are physically distinguishable by studying
  the invariant mass distribution of the lepton pairs produced. We
  then introduce an extension of the parametrization of universal
  effects in terms of oblique parameters obtained by linearly
  expanding the self-energies of the electroweak gauge-bosons to
  ${\cal O}(q^6)$. It contains eleven oblique parameters of which only
  eight are generated within SMEFT at dimension-eight: $\Shat$,
  $\That$, $W$, $Y$, $\Uhat$, $X$, plus two additional which we label
  $W'$ and $Y'$ and show how they match at linear order with the eight
  identified combinations of operator coefficients.  We then perform a
  combined analysis of a variety of LHC data on the neutral- and
  charged-current Drell-Yan processes with the aim of constraining the eight
  combinations.  We compare and combine the LHC bounds with those from
  electroweak precision $W$ and $Z$ pole observables which can only
  provide constraints in four directions of the eight-parameter space.
  We present the results in terms of limits on the eight effective
  Wilson coefficients as well as on the eight oblique parameters.  In
  each case, we study the dependence of the derived constraints on the
  order of the expansion considered.
\end{abstract}

%%%%%%%%%%%%%%%%%%%%%%%%%%%%%%%%%%%%%%%%%%%%%%%%%%%%%%%%%%%%%%%%%%%%%%
\maketitle

%%%%%%%%%%%%%%%%%%%%%%%%%%%%%%%%%%%%%%%%%%%%%%%%%%%%%%%%%%%%%%%%%%%%%%%%%%%
\section{Introduction}

The large statistics collected by the CERN Large Hadron Collider (LHC)
in its different runs have allowed for precise tests of the Standard
Model (SM) predictions as well as searches for new physics. A
particular place in this quest is held by dilepton production, the
so-called Drell-Yan (DY) process~\cite{Drell:1970wh}, which can
proceed either via neutral-current (NC) or charged-current (CC)
\[
  p p  \to \ell^+ \ell^-    \;\;\;\hbox{ and }\;\;\; p p \to \ell^\pm \nu_\ell
\]
with $\ell =e,\mu$. Involving only leptons in the final state, this
process provides a clean environment for both experimental studies and
theoretical predictions.  The LHC experimental collaborations have
taken advantage of this to perform precision SM
tests~\cite{ATLAS:2016gic,CMS:2014jea, Walia:2018fon,
  CMS:2024ony,ATLAS:2025hhn} and searches for new resonances
~\cite{ATLAS:2019erb, CMS:2021ctt, ATLAS:2019lsy, CMS:2022krd}.
Presently there is no data that is at variance with the SM, ergo
there may exist a mass gap between the electroweak and the new
physics scales. In such a scenario, hints on the new physics can first
manifest itself through deviations from the SM predictions. In this
case, it is natural to employ Effective Field Theory (EFT) as a
model-independent approach to analyze the experimental
results.%\medskip

Under the minimal assumption  that the scalar particle observed in
2012~\cite{Aad:2012tfa, ATLAS:2017jag} is, in fact,  part of an electroweak
doublet, 
the $SU(2)_L \otimes U(1)_Y$ symmetry can be linearly realized and 
the resulting EFT is the 
so-called Standard Model EFT (SMEFT). In this framework, deviations from
the SM predictions are parametrized as higher order operators
\begin{equation}
  {\cal L}_{\rm eff}  = {\cal L}_{\rm SM} + \sum_{n>4,j}
  \frac{f_{n,j}}{\Lambda^{n-4}} {\cal O}_{n,j} \;, 
\end{equation}
where $\Lambda$ is a characteristic energy scale and ${\cal O}_{n,j}$
are higher dimension operators. At the LHC the first sizable
contributions are of dimension six, i.e.
${\cal O}(\Lambda^{-2})$. It is well known that there are 59
independent dimension–six operators~\cite{Grzadkowski:2010es} leading
to 2499 arbitrary Wilson coefficients when flavor is taken into
account~\cite{Jenkins:2013wua}. The situation becomes close to
untreatable when we consider the next order (${\cal O}(\Lambda^{-4})$)
in the expansion that exhibits 44,807 possible
operators~\cite{Henning:2015alf}. As a consequence in the most general
scenario, the number of dimension-eight operators contributing to the
present observables is prohibitively large, which precludes a complete
analysis including all effects at that order.  In this context,
Drell-Yan processes have been studied in the SMEFT framework at
${\cal O}(\Lambda^{-2})$~\cite{Dawson:2018dxp,
  Dawson:2021ofa,Farina:2016rws, Panico:2021vav, Torre:2020aiz,
  Horne:2020pot} and partially at order
${\cal O}(\Lambda^{-4})$~\cite{Alte:2018xgc, Alioli:2020kez,
  Boughezal:2021tih, Boughezal:2022nof, Allwicher:2022mcg,
  Allwicher:2022gkm, Grossi:2024tou,Corbett:2024evt,Allwicher:2024mzw} where due to the large number of
operators contributing the majority of the studies consider just one
or at most a few operators at one time. %\medskip

Identifying physically motivated hypotheses to be able to capture a
large class of beyond the standard model (BSM) theories while reducing
the number of relevant operators becomes mandatory for general
studies.  One such well-motivated hypothesis is that of {\sl
Universality}, which in brief refers to BSM scenarios where { the new
physics either dominantly couples to the bosons of the Standard Model or if it couples to fermions, couples via SM currents allowing to express the EFT exclusively in terms of bosonic operators.} 
At ${\cal O} (\Lambda^{-2})$ under the assumption of C
and P conservation, this universal SMEFT (herein USMEFT) contains 16
dimension-six independent operators ~\cite{Wells:2015uba}. Recently,
Ref.~\cite{Corbett:2024yoy} presented the basis for USMEFT at
${\cal O} (\Lambda^{-4})$ which, without imposing C or P symmetries,
contains 175 dimension-eight operators. %\medskip

In this work, we perform a complete study of the
neutral- and charged-current Drell-Yan processes  and  electroweak precision
observables (EWPO) in the framework of C and P conserving universal new physics
beyond ${\cal O}(\Lambda^{-2})$. Within the USMEFT framework
there are 17 operators that contribute to the EWPO and Drell-Yan.
They are presented both in the purely bosonic form and in the {\sl
rotated} form in which the bosonic operators with higher derivatives
are traded for fermionic operators involving the SM gauge
currents. We identify eight combinations of the 17 Wilson coefficients
which are physically distinguishable in the EWPO and Drell-Yan
analyses. Furthermore, we introduce an extension of the
parametrization of universal effects in terms of oblique parameters
obtained by linearly expanding the self-energies of the electroweak
gauge bosons to ${\cal O}(q^6)$. It contains eleven oblique parameters
of which only eight are generated within the USMEFT at
${\cal O} (\Lambda^{-4})$.%\medskip

In these studies, we compare and combine the constraints derived from the
analysis of LHC DY processes with those from the $W$ and $Z$ pole
observables which can only provide bounds in four directions of the
eight-parameter space.  We present the results in terms of constraints
on the eight testable effective Wilson coefficients as well as on the eight
oblique parameters.  In each case, we study the dependence of the
derived constraints with the order of the expansion considered.
Our results show that the present data, which favours no deviation from
the SM predictions,  can robustly constrain six of the eight
parameters while strong correlations and cancellations are still
present for three of the combinations.%\medskip

The outline of the paper is as follows. In Sec.~\ref{sec:effoper} we
present the part of the USMEFT basis up to ${\cal O} (\Lambda^{-4})$
employed in this work. This section is complemented with
Appendix~\ref{app:boson8} where the full USMEFT at
${\cal O} (\Lambda^{-4})$ basis and its properties under $C$ and $P$
are listed.  Section~\ref{sec:formalism} contains some analytic
expressions of the corrections of the Drell-Yan amplitudes as derived
in the rotated basis, which is most convenient for numerical
implementation in phenomenological studies, consistently accounting
for the effect induced by the renormalization of the SM inputs at
${\cal O} (\Lambda^{-4})$. In Section ~\ref{sec:oblique} we introduce
an extension of the oblique parameters to take into account
${\cal O}(q^6)$ contributions and this section is complemented with
Appendix~\ref{app:oblique} where we show the relation between those
eight oblique parameters and the eight effective combinations
identified in Sec.~\ref{sec:formalism}. We then proceed to make a
combined analysis of the experimental results presented in
Sec.~\ref{sec:framework}.  The quantitative results of the analysis
are presented in Sec.~\ref{sec:results}, while  Sec.~\ref{sec:summary}
contains our summary.

%%%%%%%%%%%%%%%%%%%%%%%%%%%%%%%%%%%%%%%%%%%%%%%%%%%%%%%%%%%%%%%%%%%%%%
\section{Operator basis}
\label{sec:effoper}

Within the SMEFT predictions for observables at order $1/\Lambda^4$
require evaluating the SM contributions, the interference between the
$1/\Lambda^2$ amplitude (${\cal M}^{(6)}$) with the SM amplitude, the
square of the dimension-six amplitude, as well as the interference of
the $1/\Lambda^4$ amplitude with the SM, which
we represent as:
\begin{equation}
  | M_{\rm SM}|^2 + {\cal M}_{\rm SM}^\star {\cal M}^{(6)}  + | {\cal
    M}^{(6)}|^2 + {\cal M}_{\rm SM}^\star {\cal M}^{(6,2)}+ {\cal M}_{\rm SM}^\star {\cal M}^{(8)}
  \;.
\label{eq:ampli}  
\end{equation}
${\cal M}^{(8)}$ includes amplitudes with one dimension-eight operator
coefficient while ${\cal M}^{(6,2)}$ includes the contribution of the
insertion of two dimension-six Wilson coefficients in the
amplitude. %\medskip

In order to perform our analyses we must choose a basis of independent
operators. Universal theories in the context of the SMEFT refer to BSM
models for which the low-energy effects can be parametrized in terms
of operators involving exclusively the SM bosons, herein referred to
as {\sl bosonic operators} ~\cite{Wells:2015uba}.  In the EFT
framework not all operators at a given order are
independent. {In USMEFT}, integration by parts and Bianchi identities allow for the
selection of a basis of independent operators still involving only bosonic
fields. In what follows we refer to this as the {\sl bosonic
  basis}. Generically some of these operators contain higher
derivatives of the bosonic fields. As is widely known, operators
connected by the use of the classical equations of motion (EOM) of the
SM fields lead to the same $S$--matrix elements~\cite{Politzer:1980me,
  Georgi:1991ch, Arzt:1993gz,Simma:1993ky}. While for a top-down approach the
  EOM are insufficient and one needs to use field redefinitions \cite{Criado:2018sdb}, 
  this problem does not affect bottom-up EFTs as at each order the basis of operators is complete. Thus, it is possible to
trade those bosonic operators with higher derivatives for operators
involving fermions in the form of combinations of SM currents herein
called {\sl fermionic operators}. We will refer to this basis as the
{\sl rotated basis}.

%%%%%%%%%%%%%%%%%%%%%%%%%%%%%%%%%%%%%%%%%%%%%%%%%%%%%%%
\subsection{Dimension-six basis}
\label{sec:dim6}

The complete list of universal dimension-six operators is presented in
Ref.~\cite{Wells:2015uba}. Assuming that the fermion masses (Yukawa
couplings) are negligible as well as requiring $C$ and $P$ conservation one can
identify six independent bosonic operators contributing to the weak
boson propagators, hence contributing to EWPO and/or Drell-Yan
processes, of which two can be rotated into fermionic operators by the
EOM. Thus we have the bosonic and rotated basis of operators relevant
for Drell-Yan listed in Table~\ref{tab:uniopd6}. %\medskip

%%%%%%%%%%%%%%%%%%%%%%%%%%%%%%%%%%%%%%%%%%%%%%%%%%%%%%%
\begin{table}
\begin{tabular}{|l||l|l||l||l|l|}
  \hline
  \multicolumn{2}{|c|}{{Bosonic Basis}} &
 \text{Coefficient} &
 \multicolumn{2}{|c|}{{Rotated Basis}}&
\text{Coefficient} 
 \\\hline
$\qpone$
& $(D_\mu H^\dagger H)(H^\dagger D^\mu H)$
& $b_{\Phi,1}$
&$\qpone$&  $(D_\mu H^\dagger H)(H^\dagger D^\mu H)$ &
$c_{\Phi,1}$\\
$\qww$
& $H^{\dagger} {W^I}_{\mu \nu} {W}^{I,\mu \nu} H$ &
$b_{WW}$
&$\qww$ &$H^{\dagger} {W^I}_{\mu \nu} {W}^{I,\mu \nu} H$ &
$c_{WW}$\\ 
$\qbb$ & $H^{\dagger} {B}_{\mu \nu} {B}^{\mu \nu} H$
& $b_{BB}$
&$\qbb$ &  $H^{\dagger} {B}_{\mu \nu} {B}^{\mu \nu} H$
&$c_{BB}$\\
$\qbw$
&  $H^\dagger {B}_{\mu\nu}\tau^I {W}^{I, \mu\nu} H$ &
$b_{BW}$
&$\qbw$ &$H^\dagger {B}_{\mu\nu}\tau^I {W}^{I, \mu\nu} H$ &
${c_{BW}}$\\\hline
${\cal R}_{2W}$& $-\frac{1}{2} (D^\nu W_{\mu\nu}^I)^2$ &
$r_{2W}$
&  $Q_{2JW}$ &  $J^I_{W\mu} J^{I\mu}_W$ &
$c_{2JW}$\\
${\cal R}_{2B}$& $-\frac{1}{2} (D^\nu B_{\mu\nu})^2$ & $r_{2B}$ & 
$Q_{2JB}$ &  $J_{B\mu} J^\mu_B$ & $c_{2JB}$\\
\hline
\end{tabular}
\caption{ C and P conserving dimension-six operators for universal
  theories and their respective Wilson coefficients. $H$ stands for
  the SM Higgs doublet and ${W}_{\mu\nu}^I $ and ${B}_{\mu\nu}$ are the
  $SU(2)_L$ and $U(1)_Y$ field strength tensors respectively.
  $\tau^I$ stands for the Pauli matrices.}
\label{tab:uniopd6}
\end{table}
%%%%%%%%%%%%%%%%%%%%%%%%%%%%%%%%%%%%%%%%%%%%%%%%%%%%%%%

In the rotated basis the operators involve the SM fermion currents 
\begin{eqnarray}
  J_B^\mu &= &g' \displaystyle\sum_{f \in \{q,l,u,d,e\}} \sum_a Y_f
    \bar{f}_a\gamma^\mu f_a\; , \nonumber \\
  J^{I\mu}_{W} &=& \frac{g}{2} \displaystyle\sum_{f \in \{q,l\}}
  \sum_a \bar{f}_a\gamma^\mu \tau^I f_a\;
\end{eqnarray}
with $Y_f$ standing for the fermion $f$ hypercharge, $q$ and $l$ are
the quark and lepton doublets and $u$, $d$ and $e$ represent the
fermion singlets, and the sum over $a$ is over generations. The $SU(2)_L$ and $U(1)_Y$ gauge couplings are $g$
and $g^\prime$. %\medskip

As mentioned above, the coefficients in both bases are related by EOM
as:
\begin{equation}
  \begin{array}{l}
    c_{\Phi,1}=b_{\Phi,1}+\frac{g'^2}{2}r_{2B} \;,\\
    c_{WW} = b_{WW} - \frac{g^2}{4}r_{2W} \;,\\
    c_{BB} = b_{BB} - \frac{g'^2}{4}r_{2B} \;,\\
    c_{BW} = b_{BW} - \frac{gg'}{4}(r_{2B}+r_{2W}) \;,\\
    c_{2JW}= -\frac{1}{2}r_{2W} \;,\\
    c_{2JB}= -\frac{1}{2}r_{2B} \;.\\
  \end{array}
 \label{eq:rot6}   
\end{equation}
These relations are derived by identifying operator relations from the EOM and reducing the set of equations.

%%%%%%%%%%%%%%%%%%%%%%%%%%%%%%%%%%%%%%%%%%%%%%%%%%%%%%%
\subsection{Dimension-eight operators}
\label{sec:dim8}

The full basis of dimension-eight operators for universal theories was
presented in Ref.~\cite{Corbett:2024yoy} and contains 175 operators.
It consists of 89 bosonic operators already included in Murphy's
basis~\cite{Murphy:2020rsh} and 86 additional bosonic operators with
higher derivatives which can be rotated into fermionic operators by
the EOM as shown in Ref.~\cite{Corbett:2024yoy}.  We list in
Appendix~\ref{app:boson8} the full list of 175 operators in the
bosonic basis together with their $C$ and $P$ properties.  From those
we identify 11 independent bosonic operators contributing to the weak
boson propagators of which 8 can be rotated into combinations
involving 10 fermionic operators by the EOM. We list the final 
bosonic and rotated basis of independent operators relevant for Drell-Yan in
Table~\ref{tab:uniopd8}.
%
%%%%%%%%%%%%%%%%%%%%%%%%%%%%%%%%%%%%%%%%%%%%%%%%%%%%%%%
\begin{table}
\begin{tabular}{|l||l|l||l||l|l|}
\hline
\multicolumn{2}{|c|}{{Bosonic Basis}} & {Coeff} & 
\multicolumn{2}{|c|}{{Rotated Basis}} & {Coeff}  \\\hline
$Q^{(2)}_{H^6}$ & $(H^\dagger H) (H^\dagger\sigma^I H) (D_\mu H)^\dagger \sigma^I D^\mu H$ & $b^{(2)}_{H^6}  $ & 
$Q^{(2)}_{H^6} $ & $ (H^\dagger H) (H^\dagger
\sigma^I H) (D_\mu H)^\dagger \sigma^I D^\mu H
$ & $ c^{(2)}_{H^6}$\\[+0.1cm]
$Q_{WBH^4}^{(1)}$ & $  (H^\dag H) (H^\dag \tau^I H) W^I_{\mu\nu} B^{\mu\nu}
$ & $ b^{(1)}_{BWH^4}  $ & 
$Q_{WBH^4}^{(1)} $ & $  (H^\dag H) (H^\dag \tau^I H) W^I_{\mu\nu} B^{\mu\nu}
$ & $ c^{(1)}_{BWH^4} $ \\[+0.1cm]
$Q_{W^2H^4}^{(3)}
$ & $ (H^\dag \tau^I H) (H^\dag \tau^J H)W^I_{\mu\nu} W^{J\mu\nu}  
$ & $ b^{(3)}_{W^2H^4} $ & 
$Q_{W^2H^4}^{(3)}
$ & $ (H^\dag \tau^I H) (H^\dag \tau^J H)W^I_{\mu\nu} W^{J\mu\nu}  
$ & $ c^{(3)}_{W^2H^4} $ \\[+0.1cm]
\hline
$R^{(1)}_{B^2D^4} $ & $  D^\rho D^{\alpha}B_{\alpha\mu} D_\rho D_{\beta}B^{\beta\mu}
$ & $ r^{(1)}_{B^2D^4} $ & $
  Q^{(2)}_{\psi^4 D^2} $ & $
  D^\alpha  J_B^\mu D_\alpha J_{B\mu}
$ & $c^{(2)}_{\psi^4 D^2}$
  \\[+0.1cm]
 &  &  & 
$Q^{(1)}_{\psi^2 H^2D^3} $ & $
\begin{array}{l}    i\,(D^\mu J_B^{\nu}+D^\nu J_B^{\mu}) \\
  \times (D_{(\mu}D_{\nu)} H^\dagger H-H^\dagger D_{(\mu}D_{\nu)} H)
\end{array}
  $ & $\begin{array}{l}   
    c^{(1)}_{\psi^2H^2 D^3}%=-\frac{g'}{2} r^{(1)}_{B^2 D^4}
    \\
%   = -\frac{g'}{2} c^{(2)}_{\psi^4 D^2}
  \end{array}$\\[+0.4cm]
$R^{(1)}_{W^2D^4} $ & $ D^\rho D^{\alpha}W^I_{\alpha\mu} D_\rho D^{\beta}W_{\beta}^{I,\mu}
$ & $ r^{(1)}_{W^2D^4} $ & $
   Q^{(3)}_{\psi^4 D^2} $ & $  D^\alpha J_W^{I\mu} D_\alpha J_{W\mu}^I
 $ & $ c^{(3)}_{\psi^4 D^2}$\\[+0.1cm]
   &  &  & $
  Q^{(2)}_{\psi^2 H^2D^3} $ & $
  \begin{array}{l}
      i\,(D^\mu J_W^{I\nu}+D^\nu J_W^{I\mu}) \\\times (D_{(\mu}D_{\nu)}
  H^\dagger\tau^I H-H^\dagger\tau^I D_{(\mu}D_{\nu)}H)
\end{array}
   $ & $ \begin{array}{l}
     c^{(2)}_{\psi^2H^2 D^3}%=-\frac{g}{2} r^{(1)}_{W^2 D^4}
    \\
%    =-\frac{g}{2} c^{(3)}_{\psi^4 D^2}
  \end{array}$\\[+0.4cm]
$R^{(1)}_{WH^4D^2} $ & $ i (D^\mu W^I_{\mu\nu}) (H^\dag \overleftrightarrow{D}^{I \nu} H)  (H^\dag H) 
$ & $ r^{(1)}_{W H^4D^2} $ & $
Q^{(1)}_{\psi^2 H^4 D} $ & $
i\,J_B^\mu (H^\dagger\overleftrightarrow{D}_{\mu} H) (H^\dagger H)
$ & $c^{(1)}_{\psi^2 H^4 D} $\\ [+0.1cm]            
$R^{\prime (2)}_{WH^4D^2} $ & $ \epsilon^{IJK}(H^\dag \tau^I H) D^{\nu}(H^\dag \tau^{J} H) (D^{\mu}W^K_{\mu\nu})
$ & $ r^{\prime(2)}_{W H^4D^2} $ & $
Q^{(2)}_{\psi^2 H^4 D} $ & $   \begin{array}{r}i\,J_W^{I\mu}\left[ (H^\dagger\overleftrightarrow{D}^I_{\mu} H) (H^\dagger H)\right.\\\left.+(H^\dagger\overleftrightarrow{D}_\mu H) (H^\dagger \tau^I H)\right]\end{array}          
$ & $c^{(2)}_{\psi^2 H^4 D}$ \\[+0.1cm]
$R^{(1)}_{BH^4D^2} $ & $  i (D_{\alpha}B^{\alpha \mu})(H^\dagger \overleftrightarrow{D}_{\mu} H) (H^\dagger H)
$ & $ r^{(1)}_{BH^4D^2} $ & $
Q^{(4)}_{\psi^2 H^4 D} $ & $
\epsilon^{IJK}\,J_W^{I\mu}(H^\dagger\tau^J_{\mu} H) D_\mu(H^\dagger \tau^K H)
$ & $c^{(4)}_{\psi^2 H^4 D}$\\         [+0.1cm]    
$R^{(9)}_{B^2H^2 D^2} $ & $ (D^\mu B_{\mu\alpha})(D_\nu  B^{\nu\alpha})  (H^\dagger H)
$ & $ r^{(9)}_{B^2H^2D^2}
$ & $
Q^{(4)}_{\psi^4 H^2} $ & $
 J_B^{\mu}J_{B\mu}   (H^\dagger H)
$ & $c^{(4)}_{\psi^4 H^2} $\\                 [+0.1cm]
$R^{(9)}_{W^2H^2 D^2} $ & $ (D^\mu W^I_{\mu\alpha}) (D_\nu W^{I,\nu\alpha})  (H^\dagger H) 
$ & $ r^{(9)}_{W^2H^2D^2} $ & $
Q^{(5)}_{\psi^4 H^2} $ & $
 J_W^{I\mu}J_{W\mu}^I  (H^\dagger H)
$ & $c^{(5)}_{\psi^4 H^2} $\\                 [+0.1cm]
$R^{(13)}_{BWH^2 D^2} $ & $ (D^\mu B_{\mu\alpha})(D_\nu  W^{I,\nu\alpha})  (H^\dagger\tau^I H)
$ & $ r^{(13)}_{BW H^2D^2}
$ & $
Q^{(7)}_{\psi^4 H^2} $ & $
J_W^{I\mu}J_{B\mu} (H^\dagger\tau^I H)
$ & $c^{(7)}_{\psi^4 H^2} $\\                 
\hline
\end{tabular}
\caption{Independent dimension-eight  C and P conserving 
USMEFT operators relevant for Drell-Yan.}
\label{tab:uniopd8}
\end{table}
%%%%%%%%%%%%%%%%%%%%%%%%%%%%%%%%%%%%%%%%%%%%%%%%%%%%%%%
%
The coefficients of the two bases are related by the EOM as:
\begin{equation}
  \begin{array}{l}
   c^{(2)}_{H^6} = b^{(2)}_{H^6} +
    \frac{g^2g'^2}{4}r^{(1)}_{B^2D^4}+g^2g'^2r^{(1)}_{W^2D^4} +
    \frac{g'^2}{2}r^{(9)}_{B^2H^2D^2}+\frac{gg'}{2}r^{(13)}_{BWH^2D^2}
    - g' r^{(1)}_{BH^4D^2} + g r'^{(2)}_{WH^4D2} \;, \\[3pt]
   c^{(1)}_{BWH^4} = b^{(1)}_{BWH^4} -\frac{g^3g'}{2}r^{(1)}_{W^2D^4} \;,\\[3pt]
   c^{(3)}_{W^2H^4} = b^{(3)}_{W^2H^4}\\[3pt]
   c^{(1)}_{\psi^2H^4D}={ g'g^2 r^{(1)}_{W^2D^4}}-
    r^{(1)}_{BH^4D^2}+g'r^{(9)}_{B^2H^2D^2}+\frac{g}{2}
    r^{(13)}_{BWH^2D^2} \;,\\[3pt]
   c^{(2)}_{\psi^2H^4D}={\frac{g'^2g}{4} r^{(1)}_{B^2D^4}}+\frac{g^3}{4} r^{(1)}_{W^2D^4}-\frac{1}{2} r^{(1)}_{WH^4D^2} 
  +\frac{g}{2} r^{(9)}_{W^2H^2D^2}+{ \frac{g'}{4} r^{(13)}_{BWH^2D^2}}
    \;,\\[3pt]
   c^{(4)}_{\psi^2H^4D}=\frac{g^3}{4} r^{(1)}_{W^2D^4} -\frac{1}{2} r^{(1)}_{WH^4D^2} - r^{\prime (2)}_{WH^4D^2} 
   +\frac{g}{2} r^{(9)}_{W^2H^2D^2}
   { -\frac{g'}{4} r^{(13)}_{BWH^2D^2}} \;,\\[3pt]
   c^{(4)}_{\psi^4H^2}
   =r^{(9)}_{B^2H^2D^2} \;,\\[3pt]
   c^{(5)}_{\psi^4H^2}=r^{(9)}_{W^2H^2D^2} \;,\\[3pt]
   c^{(7)}_{\psi^4H^2}=r^{(13)}_{BWH^2D^2} \;,\\[3pt]
   c^{(2)}_{\psi^4 D^2}
   =r^{(1)}_{B^2D^4} \;,\\[3pt]
   c^{(3)}_{\psi^4 D^2}
   =r^{(1)}_{W^2D^4} \;.\\[3pt]
  \end{array}
 \label{eq:rot8} 
\end{equation}
In addition the EOM imply two relations connecting the coefficients of
the following operators in the rotated basis
\begin{equation}
c^{(1)}_{\psi^2 H^2 D^3}=-\frac{g'}{2} c^{(2)}_{\psi^4 D^2}  \;, \;\;\;\;
c^{(2)}_{\psi^2 H^2 D^3}=-\frac{g}{2} c^{(3)}_{\psi^4 D^2}\;.
\label{eq:link}
\end{equation}

%%%%%%%%%%%%%%%%%%%%%%%%%%%%%%%%%%%%%%%%%%%%%%%%%%%%%%%%%%%%%%%%%%%%%%
\section{Corrections to the Drell-Yan amplitudes}
\label{sec:formalism}

In total, we have identified six dimension-six and eleven
dimension-eight operators entering the Drell-Yan production in
USMEFT. As mentioned above, one can choose to work with the fully
bosonic basis of operators containing some with higher derivatives of
the bosonic fields, or with the rotated basis in which these operators
have been rotated into fermionic operators. In order to consistently
include the new effects into realistic simulations of  the experimental
observables up to order ${\cal O}(\Lambda^{-4})$, we numerically evaluate
the corresponding event rates with standard numerical tools:  
MadGraph5\_aMC@NLO~\cite{Frederix:2018nkq} with UFO files generated
with FeynRules~\cite{Christensen:2008py, Alloul:2013bka} including all
relevant operators. For this purpose it is more convenient to work with the
operators in the rotated basis  including in addition the indirect effects
induced by the finite renormalization of the SM
parameters~\cite{Corbett:2023qtg}. %\medskip

In this work, we adopt as input parameters
$\{ \widehat{\alpha}_{\rm em} \;,\; \widehat{G}_F \;,\;
\widehat{M}_Z\}$ and consider the following three relations to define
the renormalized parameters
\begin{eqnarray}
  &&\hat{e} = \sqrt{ 4 \pi \widehat{\alpha}_{\rm em}} \;,
     \nonumber
  \\
  && \hat{v}^2 = \frac{1}{\sqrt{2} \, \widehat{G}_F} \;,
     \label{eq:vhat}
  \\
  &&\hat{c}^2 \hat{s}^2 = \frac{\pi \widehat{\alpha}_{\rm em}}{\sqrt{2}\,
     \widehat{G}_F \widehat{M}^2_Z} \;,
     \nonumber
\end{eqnarray}
where $\sh$ ($\ch$) is the sine (cosine) of the weak mixing angle
$\hat{\theta}$. %\medskip

For convenience we parametrize the contributions of fermionic
operators to the muon decay width as
\begin{equation}
  \left[2\langle H^\dagger H \rangle - \frac{1}{\sqrt{2} \widehat{G}_F}
  \right]_{\rm fermionic}\equiv  \frac{\vh^4}{\Lambda^2} \Delta_{4F}
  + \frac{\hat{v}^6}{\Lambda^4} \Delta^{(8)}_{4F}  \;,
\label{eq:vt}
\end{equation}
where $\Delta_{4F}$ ($\Delta_{4F}^{(8)}$) contains the dimension-six
(-eight) contributions. In USMEFT there is just one fermionic
dimension-six operator that contributes:
\begin{equation}
  \Delta_{4F}
  =-\frac{\eh^2}{2\sh^2} c_{2JW}  \;.
\end{equation}
%

%%%%%%%%%%%%%%%%%%%%%%%%%%%%%%%%%%%%%%%%%%%%%%%%%%%%%%%
\subsection{$Z$ and $W$ couplings}

After finite renormalization of the SM inputs and accounting for the
direct contribution from the fermionic dimension-eight operators
containing two fermion fields we can parametrize the $Z$ coupling to
fermion pairs $\bar{f}f$ as
\begin{equation}
  \frac{\eh}{\sh\ch} ~\left[\hat{g}_{L,R}^f\, \left(1+\Delta \overline{g}_1
    +\Delta g_1^\square+\frac{p^2}{\widehat{M}_Z^2}\Delta g_1'\right)
    +Q^f\, \left(\Delta \overline{g}_2
    +\Delta g_2^\square+\frac{p^2}{\widehat{M}_Z^2}\Delta g_2'\right)\right]\;,
\label{eq:dgZ}
\end{equation}
and the $W$ coupling to left-handed fermions as
\begin{equation}  
 \frac{1}{\sqrt{2}} \frac{\eh}{\sh} ~
\left(1+\Delta \overline{g}_W
    +\Delta g_W^\square+\frac{p^2}{\widehat{M}_W^2}\Delta g_W'\right)\;,
\label{eq:dgW}  
\end{equation}
where $\hat g^f_L=T_3^f-\sh^2 Q^f$, $\hat g^f_R=-\sh^2 Q^f$, $T_3^f$
is the fermion's third component of isospin, and $Q^f$ is its
charge. In the expressions above $p^2$ is the square of the
four-momentum of the corresponding gauge boson. %\medskip

The $\Delta \overline{g}_{1,2,W}$ pieces contain corrections which are
dimension-six at leading order with additional contributions from
either ${Q_{WW}}$ and ${Q_{BB}}$\footnote{The coefficients of operators ${ Q}_{BB}$ and ${} Q_{WW}$
  induce an overall renormalization of the $W^I$ and $B$ field
  wave functions that can be absorbed by a redefinition of the gauge
  boson coupling constants at all orders. However, this does not apply
  to dimension-six operators that involve powers of the gauge
  couplings without corresponding powers of the $W^I$ and/or $B$ gauge
  fields. In particular, for ${ Q}_{2JW}$, ${ Q}_{2JB}$ and
  ${ Q}_{BW}$ the field redefinitions give rise to
  ${\cal O}(\Lambda^{-4})$ terms proportional to $c_{2JW} c_{WW}$,
  $c_{2JB} c_{BB }$ and $c_{WB} (c_{WW}+c_{BB})$ respectively.}
or dimension-eight operators which are only resolvable with Higgs
observables.  They read
\begin{eqnarray}
  \Delta \overline{g}_1= &&
  - \frac{1}{4} 
  \left[ 2 \overline{\Delta}_{4F} 
    + \overline{c}_{\Phi,1} \right] \frac{\vh^2}{\Lambda^2}
 \label{eq:dg1b} \;, \\
\Delta \overline{g}_2=
&&-
\frac{\sh_2}{8\ch_2} \Big[
  \sh_2 \left(2 \overline{\Delta}_{4F} 
  + \overline{c}_{\Phi,1}\right)
  +4
  \overline{c}_{BW} \Big] \frac{\vh^2}{\Lambda^2} \;,
\label{eq:dg2b}\\
 \Delta \overline{g}_W=&&
 -\frac{1}{4\ch_2}  \left [
      2\sh_2 \overline{c}_{BW}
        +2 \ch^2 \overline{\Delta}_{4F}
        +\ch^2 \overline{c}_{\Phi,1} \right]\frac{\vh^2}{\Lambda^2}
 -\frac{1}{2}   \overline{c}^{(3)}_{W^2H^4} \frac{\vh^4}{\Lambda^4} \;,
\label{eq:dgwb}
\end{eqnarray}
with $\hat{c}_{n}=\cos (n\hat{\theta})$ and
$\hat{s}_{n}=\sin (n\hat{\theta})$. In addition we have introduced the
effective coupling combinations
\begin{eqnarray}
  \overline{\Delta}_{4F} &=&-\frac{\eh^2}{2\sh^2} c_{2JW}
(1{-2\frac{\vh^2}{\Lambda^2} c_{WW}})
  -\frac{\hat e^2}{4\sh^2} c^{(5)}_{\psi^4 H^2}\frac{\vh^2}{\Lambda^2}
\;, 
  \label{eq:d4fb}\\
  \overline{c}_{BW}
 &=&
  c_{BW} \big(1-{ \frac{\vh^2}{\Lambda^2}} (c_{WW}+c_{BB})\big) + 
  \frac{1}{2}\left[ c^{(1)}_{WBH^4} 
      +\frac{\eh}{2\sh}
      c^{(1)}_{\psi^2 H^4D} +\frac{\eh}{\ch}
      c^{(2)}_{\psi^2 H^4D}\right]\frac{\vh^2}{\Lambda^2} \;,
\label{eq:cbwb}\\
  \overline{c}_{\Phi,1}
 & =& {c}_{\Phi,1}
+\left[c_{H^6}^{(2)}-\frac{\eh}{\ch}c^{(1)}_{\psi^2 H^4D}
     -\frac{\eh}{\sh}
     \big(c^{(2)}_{\psi^2 H^4D}-c^{(4)}_{\psi^2 H^4D}\big)
     \right]\frac{\vh^2}{\Lambda^2}\;,
\label{eq:cp1b}\\
\overline{c}^{(3)}_{W^2H^4}
 & =&c^{(3)}_{W^2H^4}+\frac{\eh}{2\sh}\left(
 c^{(2)}_{\psi^2 H^4D}-c^{(4)}_{\psi^2 H^4D}\right)\;.
\label{eq:c3whb}
\end{eqnarray}
We collect in $\Delta g_{1,2}^{\square}$ the additional contributions
which are quadratic in the dimension-six Wilson coefficients:
\begin{eqnarray}
  \Delta g_1^{\square}=&&  \frac{1}{32} 
  \frac{1}{\ch_2}\left[{ 16\sh^2 (\Delta_{4F})^2}+
    3\ch_2\left( 4(\Delta_{4F})^2
    +  ( c_{\Phi,1})^2\right)+
     4  \Delta_{4F}   c_{\Phi,1}
    +16 \sh_2  \Delta_{4F}   c_{BW}
        \right] \frac{\vh^4}{\Lambda^4}  \;,
\label{eq:dg1sq}
\\
\Delta g^\square_2=
  &&
  \frac{\sh_2^2}{128\ch_2^3}
  \left[
{ 32\sh^2\ch_2 (\Delta_{4F})^2}+
    (1+3\hat{c}_4)\left(4({\Delta}_{4F})^2+ ({c}_{\Phi,1})^2\right)- 32
(c_{BW})^2\right]\frac{\vh^4}{\Lambda^4} \;
  \nonumber \\
 &&
  +\frac{\sh_2}{8\ch_2^3}
  \left[4(-1+\ch^2\sh_2^2) \Delta_{4F} {c}_{BW}
    -\sh_2^2 {c}_{\Phi,1} {c}_{BW}
    -\sh^2\sh_2 \Delta_{4F} {c}_{\Phi,1}\right]\frac{\vh^4}{\Lambda^4}
  \;,
  \label{eq:dg2sq}\\
\Delta g^\square_W= &&
\frac{1}{32\ch_2^3} \Big[
 { 4\sh^2_2\ch_2 (\Delta_{4F})^2}+ 
  \ch^4 (5 \ch_2-2) \left(4({\Delta}_{4F})^2 +({c}_{\Phi,1})^2\right) 
  +4\left(-3\ch_2^3+(\ch_2-2) \right)({c}_{BW})^2 \nonumber
\\
&& + 8 \sh^2\sh_2 (\ch_2-2) {\Delta}_{4F} {c}_{BW}
- 4 \sh^2\sh_2 (\ch_2+2) {c}_{BW} {c}_{\Phi,1}
+(\ch_2+1)\left((5\ch_2-2)
-\ch_2^2\right) {\Delta}_{4F} {c}_{\Phi,1}\Big]
   \frac{\vh^4}{\Lambda^4} \;.
  \label{eq:dgwsq}
\end{eqnarray}
Furthermore $\Delta g'_{1,2,W}$ contain the coefficients of momentum
dependent corrections to the gauge boson couplings arising from
$Q^{(1),(2)}_{\psi^2 H^2D^3}$. Their coefficients are related with
those of $\psi^4 D^2$ operators by the universality condition as
Eq.~\eqref{eq:link} so
\begin{align}
  \Delta g^\prime_1&=-\frac{\eh^3}{4\ch^2\sh^2}
\left({ \frac{1}{\ch} c^{(1)}_{\psi^2H^2D^3} + \frac{1}{\sh} c^{(2)}_{\psi^2H^2D^3} }\right)
\frac{\vh^4}{\Lambda^4}
&&=\frac{\eh^4}{8\ch^4\sh^4}
\left( \sh^2 c^{(2)}_{\psi^4D^2} + \ch^2 c^{(3)}_{\psi^4D^2}\right)
\frac{\vh^4}{\Lambda^4}
\;,
\label{eq:dg1p}
\\
\Delta g^\prime_2&=
-\frac{\eh^3}{{ 4\ch^2\sh^2}}
\left( \sh c^{(2)}_{\psi^2H^2D^3} -\ch c^{{ (1)}}_{\psi^2H^2D^3}\right)\frac{\vh^4}{\Lambda^4}
&&=
\frac{\eh^4}{{ 8\ch^2\sh^2}}
\left( c^{(3)}_{\psi^4D^2} -c^{(2)}_{\psi^4D^2}\right)\frac{\vh^4}{\Lambda^4}\;,
\label{eq:dg2p}\\
\Delta g^\prime_W&=
-\frac{{ \eh^3}}{4\sh^3}
c^{(2)}_{\psi^2H^2D^3} \frac{\vh^4}{\Lambda^4}
&&=
\frac{\eh^4}{8\sh^4}
c^{(3)}_{\psi^4D^2} \frac{\vh^4}{\Lambda^4} \;.
\label{eq:dgwp}
\end{align}
Lastly  the $W$ mass  correction is:
\begin{eqnarray}
  \frac{\Delta  M_W}{\widehat M_W} =
        && -\frac{1}{4\ch_2} \frac{\vh^2}{\Lambda^2} \left [
      2\sh_2 \overline{c}_{BW} 
      + 2 \sh^2\overline{\Delta}_{4F}   +\ch^2\overline{c}_{\Phi,1}
      \right]-\frac{1}{2} \frac{\vh^4}{\Lambda^4}
           \overline{c}^{(3)}_{W^2H^4} \; + \frac{(\Delta
           M_W)^\square}{\widehat M_W}\;,
           \nonumber\\
\frac{(\Delta  M_W)^\square}{\widehat M_W} && =\frac{1}{32\ch^3_2}\frac{\vh^4}{\Lambda^4}
           \Big[ { 16\sh^4\ch_2 (\Delta_{4F})^2}  - 4 \sh^4(3 \ch_2+2) ( \Delta_{4F})^2
           + \ch^4 (5\ch_2-2) (c_{\Phi,1})^2
           +4\left(-3 \ch_2^3+(\ch_2-2)\right)  (c_{BW})^2
\nonumber
\\
&& - 4 \sh^2\sh_2 (\ch_2+2) {c}_{BW} {c}_{\Phi,1}
    -4\ch^2 (7-19\ch^2+14\ch^4) \Delta_{4F} {c}_{\Phi,1}    
     -8\sh_2(6-17\ch^2+14\ch^4) \Delta_{4F} {c}_{BW}\Big]     \;,
  \label{eq:dmwt}
\end{eqnarray}
where $\widehat M_W=\frac{\displaystyle \eh \vh}{\displaystyle 2\sh}$.

%%%%%%%%%%%%%%%%%%%%%%%%%%%%%%%%%%%%%%%%%%%%%%%%%%%%%%%
\subsection{Four-fermion contact amplitudes}

In addition to the corrections to the couplings of $Z$ and $W$ bosons
to fermion pairs, there are seven contact contributions to
four-fermion amplitudes in the rotated basis:
\begin{itemize}

\item two at dimension six: $c_{2JW}$  and $c_{2JB}$ \;,

\item five at dimension eight : $c^{(2)}_{\psi^4 D^2}$ ,
  $c^{(3)}_{\psi^4 D^2}$, $c^{(4)}_{\psi^4 H^2}$,
  $c^{(5)}_{\psi^4 H^2}$, and $c^{(7)}_{\psi^4 H^2}$.

\end{itemize}
However, in Drell-Yan processes the contribution of $c_{2JB}$ and
$c^{(4)}_{\psi^4 H^2}$ ($c_{2JW}$ and $c^{(5)}_{\psi^4 H^2}$) always
enter together in the same combination. Furthermore, as before, the
dimension-six Wilson coefficients $c_{BB}$ and $c_{WW}$ induce a shift
on the gauge coupling constants which results in an associated shift
on the four-fermion operator coefficients.  Hence, we find that the
four-fermion contact amplitudes depend upon the two combinations:
\begin{eqnarray}
  \overline{c}_{2JW}
  &\equiv& c_{2JW}(1{-2\frac{\vh^2}{\Lambda^2} c_{WW}})
  + { c^{(5)}_{\psi^4 H^2} \frac{\vh^2}{2 \Lambda^2}}
  =-\frac{2\sh^2}{\eh^2}\overline{\Delta}_{4F} \;,
\label{eq:c2jwb}
  \\
  \overline{c}_{2JB}&\equiv& c_{2JB}(1-2\frac{\vh^2}{\Lambda^2}c_{BB})
  + { c^{(4)}_{\psi^4 H^2}\frac{\vh^2}{2\Lambda^2}} \;.
\label{eq:c2jbb}  
\end{eqnarray}
In addition $Q^{(7)}_{\psi^4 H^2}$ contributes to a different
combination of the momentum independent four-fermion contact
amplitudes.  Moreover, $Q^{(2)}_{\psi^4 D^2}$ and
$Q^{(3)}_{\psi^4 D^2}$ generate distinctive momentum-dependent
four-fermion vertices. %\medskip

Altogether we can parametrize the four-fermion contact amplitudes in
Drell-Yan NC and CC processes with $p^2$ momentum transfer as
\begin{eqnarray}
  {\cal M}_\text{Cont}^\text{DY,NC} &=& -\hat e^2 \frac{1}{\widehat{M}_Z^2}
  \left[{ \frac{1}{\sh^2\ch^2}}(j_Z^f)^\mu (j_Z^{f'})_\mu \Big(\overline{\cal N}_{ZZ} +{\cal N}^\square_{ZZ}
 + \frac{p^2}{\widehat{M}_Z^2} {\cal N}'_{ZZ}\Big) \right. \nonumber \\ 
 &&  + { \frac{1}{\sh\ch}}\left((j_Z^f)^\mu (j_Q^{f'})_\mu+ (j_Q^{f})^\mu (j_Z^{f'})_\mu\right)
    \Big( \overline{\cal N}_{\gamma Z}+ {\cal N}^\square_{\gamma Z} +
     \frac{p^2}{\widehat{M}_Z^2}
    {\cal N}'_{\gamma Z}\Big)  \label{eq:4FNC}\\
&& \left.   +(j_Q^f)^\mu (j_Q^{f'})_\mu\Big(\overline{\cal N}_{\gamma\gamma}
+{\cal N}^\square_{\gamma\gamma} + \frac{p^2}{\widehat{M}_Z^2}
    {\cal N}'_{\gamma\gamma}\Big)
    \right] \;,\nonumber \\
{\cal M}_\text{Cont}^{\text DY,CC} &=& -\frac{\hat e^2}{2\sh^2}\frac{1}{\widehat{M}_W^2} 
  (j_W^f)^\mu (j_W^{f'})_\mu \Big(\overline{\cal N}_{WW}
  +{\cal N}^\square_{WW} + \frac{p^2}{\widehat{M}_W^2} {\cal N}'_{WW}\Big) \;,\label{eq:4FCC}
\end{eqnarray}
where we have made use of the following currents:
\begin{eqnarray}
(j^f_Z)^\mu&=&\hat{g}^f_L\bar f_L\gamma^\mu f_L +\hat{g}^f_R\bar f\gamma^\mu f_R\, ,\\
(j^f_Q)^\mu&=&Q^f \bar f\gamma^\mu f\, ,\\
(j^f_W)^\mu&=&f_{uL} \gamma^\mu f_{dL}\, .
\end{eqnarray}
Keeping our conventions as
before, we defined the $\overline{\cal N}$ pieces containing corrections
which are dimension-six at leading order with additional irresolvable
contributions
\begin{eqnarray}
  \overline{\cal N}_{\gamma\gamma} &&= -\frac{\eh^2}{2\sh^2\ch^2}
  \left( \sh^2 \overline{c}_{2JW} + \ch^2 \overline{c}_{2JB}\right)
\frac{\vh^2}{\Lambda^2}
+ \frac{\eh^2}{4 \sh\ch} c^{(7)}_{\psi^4H^2} \frac{\vh^4}{2\Lambda^4} \;,
 \label{eq:NGG}\\
\overline{\cal N}_{ ZZ} 
&&=-\frac{\eh^2 }{2\sh^2\ch^2}\left( \ch^2 \overline{c}_{2JW}
+ \sh^2 \overline{c}_{2JB} \right)\frac{\vh^2}{\Lambda^2}-
\frac{\eh^2}{4\sh\ch } c^{(7)}_{\psi^4H^2} \frac{\vh^4}{\Lambda^4} \;,
\label{eq:NZZ} \\
 \overline{\cal N}_{\gamma Z} 
 &&= -\frac{\eh^2 }{2\sh\ch}\left( \overline{c}_{2JW} - \overline{c}_{2JB}\right)\frac{\vh^2}{\Lambda^2}
 + \ch_2 \frac{\eh^2}{8 \sh^2\ch^2}c^{(7)}_{\psi^4H^2}  \frac{\vh^4}{\Lambda^4}\;,
\label{eq:NGZ} \\
  \overline{\cal N}_{ WW} &&= -\frac{\eh^2}{2\sh^2} \overline{c}_{2JW}
\frac{\vh^2}{\Lambda^2}
  \;.
\label{eq:NWW}
\end{eqnarray}
In addition ${\cal N}^\square$ contains the terms  quadratic in the Wilson
coefficients of the dimension-six operators
\begin{eqnarray}
 {\cal N}_{\gamma\gamma}^\square &&= -\frac{\eh^2}{4 \sh^2\ch^2}\frac{1}{\ch_2}\left[ \eh^2 \ch^2 (c_{2JW})^2 - c_{2JW}\left(\sh^2\ch^2 c_{\Phi,1} + 4\ch \sh^3 c_{BW} - 4\sh^2\ch_2 c_{WW}\right) \right.\\
   && \left.+ \ch^2 c_{2JB}\left(- \eh^2 c_{2JW} -4 \ch_2 c_{BB} + 4\ch\sh c_{BW} + \sh^2 c_{\Phi,1} \right)\vphantom{\eh^2}\right]
\frac{\vh^4}{\Lambda^4}
 \;,
 \\
   {\cal N}_{ ZZ}^\square &&= -\frac{\eh^2}{4\sh^2\ch^2} \frac{1}{\sh^2 \ch_2}\left[\eh^2 \ch^2 (c_{2JW})^2 - \sh^2 c_{2JW} \left(\ch^2 c_{\Phi,1} + 4 \ch^2 \ch_2 c_{WW} + \eh^2\sh^2 c_{2JB} + 2\sh_2 \ch^2c_{BW} \right)
     \right.  \\
     &&\left.
      +\sh^4 c_{2JB}\left( 2\sh_2 c_{BW} + \sh^2 c_{\Phi,1} - 4 \ch_2 c_{BB}\right)\right] \frac{\vh^4}{\Lambda^4}      \; ,    
\\     
          {\cal N}_{ \gamma Z}^\square &&= \frac{\eh^2}{4\sh^2\ch^2}\frac{\ch}{\sh \ch_2}\left[ -\eh^2 \ch^2 (c_{2JW})^2 + \sh^2c_{2JW}\left(\ch^2 c_{\Phi,1} + 4 \ch_2 c_{WW}+2\sh_2c_{BW} - \eh^2 c_{2JB} \right) \right. \;,\\
 &&\left.+\sh^2 c_{2JB} \left(2\sh_2 c_{BW} + \sh^2 c_{\Phi,1} -4\ch_2c_{BB}\right) \right]\frac{\vh^4}{\Lambda^4} \;,
\\
  {\cal N}_{ WW}^\square &&= \frac{\eh^2}{4 \sh^2}\frac{1}{\sh^2 \ch_2} \left[-\eh^2 \ch^2(c_{2JW})^2 + \eh^2 \sh^2c_{2JW}(\ch^2 c_{\Phi,1} + 2 \ch_2 c_{WW} + 2 \sh_2 c_{BW}) \right] \frac{\vh^4}{\Lambda^4}\;,
\end{eqnarray}
and ${\cal N}'$ contains the coefficients of the momentum-dependent
four-fermion couplings:
\begin{align}  
  {\cal N}_{\gamma\gamma}^\prime &=-\frac{\eh^4}{8\sh^4\ch^4} (\sh^2 c^{(3)}_{\psi^4D^2}+\ch^2 c^{(2)}_{\psi^4D^2})\frac{\vh^4}{\Lambda^4}
&& \label{eq:NPGG}
  \;,\\
  {\cal N}_{ZZ}^\prime &=-\frac{\eh^4}{8\sh^4\ch^4} (\ch^2 c^{(3)}_{\psi^4D^2}+\sh^2 c^{(2)}_{\psi^4D^2})\frac{\vh^4}{\Lambda^4}
  &&=-\Delta g_1'\;, \label{eq:NPZZ}\\
{\cal N}_{\gamma Z}^\prime &=-\frac{\eh^4}{8\sh^3\ch^3} (c^{(3)}_{\psi^4D^2}-c^{(2)}_{\psi^4D^2})\frac{\vh^4}{\Lambda^4} &&=-{ \frac{1}{\ch \sh}}\Delta g_2'\;, \label{eq:NPGZ}\\
{\cal N}_{WW}^\prime &=-\frac{\eh^4}{8\sh^4} c^{(3)}_{\psi^4D^2}\frac{\vh^4}{\Lambda^4}&&=-\Delta g_W\;,
                                                                                          \label{eq:NPWW}
\end{align}
where, in the rightmost equivalence, we used Eqs.~\eqref{eq:dg1p},~\eqref{eq:dg2p}, and~\eqref{eq:dgwp}. 
% where in the rightmost equivalence we write the relation between the momentum
% dependent contact amplitudes generated by $Q^{(2),(3)}_{\psi^4D^2}$
% and the momentum dependent piece of the $Z$ and $W$ propagator
% generated by $Q^{(1),(2)}_{\psi^2HD^3}$ resulting from the
% universality condition in Eq.~\eqref{eq:link}

%\medskip

In summary, { at the linear level} neglecting quadratic ({\em i.e.} proportional to
(dimension-six)$^2$) coefficients, we have identified eight
combinations of Wilson coefficients which can be distinguished by
studying the invariant mass distribution of the lepton pairs produced
in Drell-Yan processes.  They can be chosen to be the five
combinations $\overline\Delta_{4F}$, $\overline{c}_{BW}$,
$\overline{c}_{\Phi,1}$, $\overline{c}^{3}_{W^2 H^4}$, and
$\overline{c}_{2JB}$ in Eqs.~\eqref{eq:d4fb}--~\eqref{eq:c3whb}, and
Eq.~\eqref{eq:c2jbb}, together with the coefficients of the
dimension-eight operators $c^{(7)}_{\psi^4 H^2}$,
$c^{(2)}_{\psi^4 D^2}$, and $c^{(3)}_{\psi^4 D^2}$.  In what follows
we refer to this set of variables as {\sl overline
  coefficients}.

%%%%%%%%%%%%%%%%%%%%%%%%%%%%%%%%%%%%%%%%%%%%%%%%%%%%%%%
\subsection{Corrections to the $Z$- and $W$-pole observables}

In our analyses we include the constraints from EWPO on the $Z$- and
$W$-pole.  The predictions for these observables can be obtained from
Eq.~\eqref{eq:dgZ} with $p^2=\widehat{M}_Z^2$ and ~\eqref{eq:dgW} with
$p^2=\widehat{M}_W^2$ respectively, with the quadratic pieces remaining the
same. We can conveniently write these
couplings at the linear level as
\begin{eqnarray}
  \Delta {g}^{Z\;\rm pole} _1&=&\Delta\overline{g}_1+\Delta g'_1=
  - \frac{1}{4} 
  \left[ 2 \widetilde{\Delta}_{4F} 
    + \widetilde{c}_{\Phi,1} \right] \frac{\vh^2}{\Lambda^2}
 \label{eq:dg1t} \\
\Delta {g}^{Z\;\rm pole}_2&=& \Delta\overline{g}_2+\Delta g'_2=
-
\frac{\sh_2}{8\ch_2} \Big[
  \sh_2 \left(2 \widetilde{\Delta}_{4F} 
  + \widetilde{c}_{\Phi,1}\right)
  +4
  \widetilde{c}_{BW} \Big] \frac{\vh^2}{\Lambda^2}
\label{eq:dg2t}\\
 \Delta g^{W\;\rm pole}_W&=& \Delta\overline{g}_W+\Delta g'_W=
 -\frac{1}{4\ch_2}  \left [
      2\sh_2 \widetilde{c}_{BW}
        +2 \ch^2 \widetilde{\Delta}_{4F}
        +\ch^2 \widetilde{c}_{\Phi,1} \right]\frac{\vh^2}{\Lambda^2}
 -\frac{1}{2}   \widetilde{c}^{(3)}_{W^2H^4} \frac{\vh^4}{\Lambda^4}
\label{eq:dgwt}  
\end{eqnarray}
with  
\begin{eqnarray}
  \widetilde \Delta_{4F} &=&  \overline{\Delta}_{4F}-
  \frac{\eh^4}{4\sh^4} c^{(3)}_{\psi^4D^2} { \frac{\vh^2}{\Lambda^2}}
  \nonumber \\
  \widetilde{c}_{BW}
&= &\overline{c}_{BW}
  +{ \frac{\eh^4}{8\sh^3\ch^3} \left(c^{(2)}_{\psi^4D^2}+c^{(3)}_{\psi^4D^2}\right) \frac{\vh^2}{\Lambda^2}}
  \nonumber
  \\
  \widetilde{c}_{\Phi,1}
&=&
\overline{c}_{\Phi,1}-
\frac{\eh^4}{2\sh^2\ch^4}
\left(c^{(2)}_{\psi^4D^2}+ \ch^2 c^{(3)}_{\psi^4D^2}\right) \frac{\vh^2}{\Lambda^2}
\label{eq:tildas}\\
\widetilde{c}^{(3)}_{W^2H^4}&=&
\overline{c}^{(3)}_{W^2H^4}+\frac{\eh^4}{4\sh^2\ch^2} c^{(3)}_{\psi^4D^2}\nonumber
\end{eqnarray}
while it still holds that 
\begin{equation}
  \frac{\Delta  M_W}{\widehat M_W} =
       -\frac{1}{4\ch_2} \frac{\vh^2}{\Lambda^2} \left [
      2\sh_2 \widetilde{c}_{BW} 
      + 2 \sh^2\widetilde{\Delta}_{4F}   +\ch^2\widetilde{c}_{\Phi,1}
      \right]-\frac{1}{2} \frac{\vh^4}{\Lambda^4}
       \widetilde{c}^{(3)}_{W^2H^4} \; + \frac{(\Delta  M_W)^\square}{\widehat M_W}\;.
\end{equation}
     
In summary, the corrections to the $Z$ and $W$ pole observables to
order ${\cal O}(\Lambda^{-4})$ which are dominant can be expressed in
terms of four combinations (herein refereed to as {\sl tilde
  coefficients}) $\widetilde{c}_{BW}$ ,$\widetilde{c}_{\Phi,1}$ ,
$\widetilde{c}^{(3)}_{W^2H^4}$, and $\widetilde{\Delta}_{4F}$ which
involve six of the eight {\sl overline coefficients}:
$\overline{c}_{BW}$, $\overline{c}_{\Phi,1}$,
$\overline{c}^{(3)}_{W^2H^4}$, $\overline{\Delta}_{4F}$,
$c^{(2)}_{\psi^4D^2}$, and $c^{(3)}_{\psi^4D^2}$.

%%%%%%%%%%%%%%%%%%%%%%%%%%%%%%%%%%%%%%%%%%%%%%%%%%%%%%%
\section{Parametrization in terms of oblique parameters}
\label{sec:oblique}

Corrections to electroweak observables from universal theories can be
described in terms of the so-called oblique parameters defined in
terms of the corrections to the gauge boson self-energies
~\cite{Peskin:1991sw,Barbieri:2004qk,Wells:2015uba}.  Expanding the
vacuum-polarization amplitudes as:
\begin{equation}
  \Pi_{VV''}(q^2)= \Pi_{VV'} (0)+ \Pi'_{VV''}(0) q^2+\frac{1}{2}
  \Pi^{\prime\prime}_{VV''}(0) q^4+\frac{1}{6}\Pi^{\prime\prime\prime}_{VV''}(0)
  q^6+\dots
\end{equation}
To order $q^4$ one can define 7 independent oblique parameters for the
weak gauge bosons
\begin{align}
  \Shat
  &= - \dfrac{\ch}{\sh}\Pi_{3B}(0) \;, \\
    \That &= \dfrac{1}{\widehat{M}_W^2} \Big[\Pi_{WW}(0) -
            \Pi_{33}(0)\Big] \;,\\
    \Uhat & =
            \Pi_{33}^\prime(0) - \Pi_{WW}^\prime(0) \;,\\
    V& = \dfrac{\widehat{M}_W^2}{2}\Big[\Pi_{WW}^{\dprime}(0) -
       \Pi_{33}^{\dprime}(0)\Big] \;, \\
    X &= -\dfrac{\widehat{M}_W^2}{2} \Pi^{\dprime}_{3B}(0) \;,\\
    Y &= -\dfrac{\widehat{M}_W^2}{2} \Pi^{\dprime}_{BB}(0) \;,\\
    W &= -\dfrac{\widehat{M}_W^2}{2} \Pi^{\dprime}_{33}(0) \;.
\end{align}
Within  the  dimension-six SMEFT only ${\widehat{S}}$, ${\widehat{T}}$,
$W$, and $Y$ are generated of which only three combinations enter in the
electroweak gauge boson pole observables. %\medskip

In the context of the SMEFT the parametrization of the amplitudes in
terms of oblique parameters defined as linear expansions of the
self-energies is only consistent with the operator expansion at first
order, {\em i.e.} at ${\cal O}(\Lambda^{-2})$.  Nevertheless, the
parametrization in terms of oblique parameters can be considered more
general than SMEFT as it can be applied also to other expansions like
the HEFT as well as implicitly containing higher-order terms in $1/\Lambda$.
With this in mind one can extend the formalism by expanding the
self-energies to order $q^6$.  This involves four additional
parameters which can be defined as
\begin{eqnarray}
    V^\prime &=& \dfrac{\widehat M_W^4}{6}\Big[\PropDer[WW]{\tprime} -
                 \PropDer[33]{\tprime}\Big] \;, \\
    X^\prime &=& -\dfrac{\widehat M_W^4}{6}  \PropDer[3B]{\tprime} \;, \\
    Y^\prime &=& -\dfrac{\widehat M_W^4}{6}  \PropDer[BB]{\tprime} \;, \\
    W^\prime &=& -\dfrac{\widehat M_W^4}{6}  \PropDer[33]{\tprime} \;.
\end{eqnarray}

Including the above self-energy contributions one can write the
Drell-Yan amplitudes in terms of the modified propagators for the NC
and CC interactions extending the expressions in
Ref.~\cite{Farina:2016rws} to include the additional oblique
parameters as
\begin{eqnarray}
&&\begin{pmatrix}
    \dfrac{1}{p^2-\widehat{M}_Z^2}
      + \dfrac{2\Delta g^O_1}{p^2 - \widehat{M}_Z^2}
-\dfrac{\Big(\varepsilon_{ZZ} + \dfrac{2}{\ch^2}\varepsilon_{ZZ}^\prime \Big)
}{\widehat{M}_W^2}
-\dfrac{\varepsilon_{ZZ}^\prime}{\widehat{M}_W^4}p^2 \hspace*{0.5cm}
&
\dfrac{\Delta g^O_2}{\sh\ch (p^2 - \widehat{M}_Z^2)}
-\dfrac{ \Big(\varepsilon_{Z\gamma} + \dfrac{1}{ \ch^2}\varepsilon_{Z\gamma}^\prime \Big)}{\widehat M_W^2}
-\dfrac{\varepsilon_{Z\gamma}^\prime}{\widehat{M}_W^4}p^2\\
* &
\dfrac{1}{p^2} - \dfrac{\varepsilon_{\gamma\gamma}}{\widehat{M}_W^2}-\dfrac{\varepsilon_{\gamma\gamma}^\prime}{\widehat{M}_W^4}p^2
    \end{pmatrix}\;,\\[+0.2cm]
&& \dfrac{1}{p^2-\widehat{M}_W^2}
    + \dfrac{2 \Delta g^O_W}{p^2 -\widehat M_W^2}
-\dfrac{\varepsilon_{WW} + 2 \varepsilon_{WW}^\prime}{\widehat{M}_W^2}-\dfrac{\varepsilon_{WW}^\prime}{\widehat{M}_W^4}p^2\;.
\end{eqnarray}
Written in this form one can identify the pieces of the modified
propagators exhibiting $(p^2-M_V^2)^{-1}$ as generated by the
amplitude with the standard model $V$ propagator and with modified
vertices to the fermions.  In terms of oblique parameters they read:
\begin{eqnarray}
  \Delta g^O_1 &=& \dfrac{1}{2}\Bigg[\widehat T - \Big(W +
                   \dfrac{2}{\ch^2}W^\prime\Big)+
                   \dfrac{2\sh}{\ch}\Big(X +
                   \dfrac{2}{\ch^2}X^\prime\Big)  -
                   \dfrac{\sh^2}{\ch^2}\Big(Y +
                   \dfrac{2}{\ch^2}Y^\prime\Big)\Bigg] \;,
\label{eq:dg1bos}
  \\[1em]
    \Delta g^O_2  &=& \dfrac{\sh^2}{\ch_2}\Bigg[\ch^2 \widehat T - \widehat S
                      + \sh^2\Big(W  + \dfrac{1}{\ch^2}W^\prime\Big)
                      -\dfrac{1 -2\sh^2\ch^2}{\sh\ch}\Big(X  +
                      \dfrac{1}{\ch^2}X^\prime\Big) + \ch^2 \Big(Y +
                      \dfrac{1}{\ch^2}Y^\prime\Big)\Bigg]\; ,\\[1em]
    \Delta g^O_W 
    &=& \dfrac{1}{2\ch_2}\Bigg[\ch^2 \widehat T - 2{ \sh^2} \widehat S -
        (1 - 3\sh^2)\Big(W  + 2 W^\prime\Big) + \sh^2 \Big(Y +
        \dfrac{1}{\ch^2}Y^\prime\Big)
        - 2 \sh\ch \Big(X + \dfrac{1}{\ch^2}X^\prime\Big)\Bigg]
 -\dfrac{1}{2} \widehat U + V + \dfrac{3}{2}V^\prime \;.
\end{eqnarray}
Moreover, the pieces proportional to $M_V^{-2}$ ($q^2/M_V^4$) are
linear combinations of the $W$, $Y$, $X$, and $V$ ($W'$, $Y'$, $X'$,
and $V'$) parameters,
\begin{eqnarray}           
  \varepsilon_{\gamma\gamma}^{(\prime)} &=& \sh^2 W^{(\prime)}
  + \ch^2 Y^{(\prime)} + 2 \sh \ch X^{(\prime)}\;, \label{eq:eepgg}
  \\
  \varepsilon_{ZZ}^{(\prime)} &=& \ch^2 W^{(\prime)} + \sh^2 Y^{(\prime)} - 2 \sh \ch X^{(\prime)}\;,
  \label{eq:eepzz}\\
  \varepsilon_{Z\gamma}^{(\prime)} &=& (\ch^2 -\sh^2) X^{(\prime)} + \sh \ch(W^{(\prime)} - Y^{(\prime)})\;,
  \label{eq:eepgz}\\
    \varepsilon_{WW}^{(\prime)} &=& W^{(\prime)} - V^{(\prime)}\;,
\label{eq:eepww}
\end{eqnarray}
and they are generated by the contact four-fermion operators in the
rotated basis.  Additionally, the correction to the $W$ mass is:
\begin{equation}
  \dfrac{\Delta M_W}{\widehat M_W} =\Delta g^O_W - \dfrac{1}{2} \Big(
  V + 2 V^\prime  - W - 2 W^\prime \Big) \;.
\label{eq:mwoblique} 
\end{equation}

The relation between the parametrization in terms of oblique parameters
and the USEMFT expressions in terms of coefficients of the operators
in the rotated basis can be made explicit by computing the oblique
parameters in the bosonic basis and then applying Eqs.~\eqref{eq:rot6}
and ~\eqref{eq:rot8}.  We list those expressions in
Appendix~\ref{app:oblique}.  As seen in the appendix, within USMEFT up
to dimension-eight operators, there are eight non-vanishing oblique
parameters: $\Shat$, $\That$, $\Uhat$, $W$, $Y$, $X$, $W'$, and $Y'$.
Introducing the resulting expressions of these oblique parameters in
terms of the coefficients in the rotated basis in
Eqs.~\eqref{eq:dg1bos}--~\eqref{eq:epww} one finds
\begin{align}
  \Delta g_1^O
  &=- \frac{1}{4} 
  \left[ 2 \overline{\Delta}_{4F} 
    + \overline{c}_{\Phi,1} \right] \frac{\vh^2}{\Lambda^2}
+\frac{\eh^4}{8\ch^4\sh^4}
\left( \sh^2 c^{(2)}_{\psi^4D^2} + \ch^2 c^{(3)}_{\psi^4D^2}\right)
\frac{\vh^4}{\Lambda^4}&&=\Delta\overline{g}_1+\Delta g_1'\\
\Delta g_2^O &=
\frac{\sh_2}{8\ch_2} \Big[
  \sh_2 \left(2 \overline{\Delta}_{4F} 
  + \overline{c}_{\Phi,1}\right)
  +4
  \overline{c}_{BW} \Big] \frac{\vh^2}{\Lambda^2}
+\frac{\eh^4}{{ 8\ch^2\sh^2}}
\left( c^{(3)}_{\psi^4D^2} -c^{(2)}_{\psi^4D^2}\right)\frac{\vh^4}{\Lambda^4}
&&=\Delta\overline{g}_2+\Delta g_2'\\
 \Delta g^O_W&=
 -\frac{1}{4\ch_2}  \left [
      2\sh_2 \overline{c}_{BW}
        +2 \ch^2 \overline{\Delta}_{4F}
        +\ch^2 \overline{c}_{\Phi,1} \right]\frac{\vh^2}{\Lambda^2}
 -\frac{1}{2}   \overline{c}^{(3)}_{W^2H^4} \frac{\vh^4}{\Lambda^4}
+\frac{\eh^4}{8\sh^4}
c^{(3)}_{\psi^4D^2} \frac{\vh^4}{\Lambda^4}
&&=\Delta\overline{g}_W+\Delta g_W'\\
\varepsilon_{\gamma\gamma}&=
-\frac{\eh^2}{2\sh^2}\left(\sh^2\overline{c}_{2JW}+\ch^2\overline{c}_{2BW}
\right) \frac{\vh^2}{\Lambda^2}
+{ \frac{\ch\eh^2}{4\sh}}c^{(7)}_{\psi^4H^2} \frac{\vh^4}{\Lambda^4}
&&=\ch^2 \overline{\cal N}_{\gamma\gamma}
\label{eq:epgg}
\\
\varepsilon_{ZZ}
&=-\frac{\eh^2}{2\sh^2}\left(\ch^2\overline{c}_{2JW}+\sh^2\overline{c}_{2BW}
\right) \frac{\vh^2}{\Lambda^2}
-{ \frac{\ch\eh^2}{4\sh}}c^{(7)}_{\psi^4H^2} \frac{\vh^4}{\Lambda^4}
&&= \ch^2 \overline{\cal N}_{ZZ}
\label{eq:epzz}
\\
\varepsilon_{Z\gamma}&=
-\frac{\eh^2\ch}{2\sh}\left(\overline{c}_{2JW}-\overline{c}_{2BW}\right)
\frac{\vh^2}{\Lambda^2}
+\ch_2\frac{\eh^2}{8\sh^2}c^{(7)}_{\psi^4H^2} \frac{\vh^4}{\Lambda^4}
&&= \ch^2 \overline{\cal N}_{\gamma Z}
\label{eq:epgz}
\\
\varepsilon_{WW}&=
-\frac{\eh^2}{2\sh^2}\overline{c}_{2JW}\frac{\vh^2}{\Lambda^2} &&=
\overline{\cal N}_{WW}
\label{eq:epww}
\\
\varepsilon_{\gamma\gamma}^\prime &=-\frac{\eh^4}{8\sh^4} (\sh^2 c^{(3)}_{\psi^2D^2}+\ch^2 c^{(2)}_{\psi^2D^2})\frac{\vh^4}{\Lambda^4}
&&=\ch^4 {\cal N}'_{\gamma\gamma} 
\\
\varepsilon_{ZZ}^\prime &=-\frac{\eh^4}{8\sh^4} (\ch^2 c^{(3)}_{\psi^2D^2}+\sh^2 c^{(2)}_{\psi^2D^2})\frac{\vh^4}{\Lambda^4}
&&=\ch^4 {\cal N}'_{ZZ} 
=- \ch^4\Delta g^\prime_1
\\
\varepsilon_{Z\gamma}^\prime &=-\ch\frac{\eh^4}{8\sh^3} (c^{(3)}_{\psi^2D^2}-c^{(2)}_{\psi^2D^2})\frac{\vh^4}{\Lambda^4}
&&=\ch^4 {\cal N}'_{\gamma Z} 
=- \ch^4 { \frac{1}{\ch \sh}}\Delta g^\prime_2
\\
\varepsilon_{WW}^\prime &=-\frac{\eh^4}{8\sh^4} c^{(3)}_{\psi^2D^2}\frac{\vh^4}{\Lambda^4}
&&={\cal N}'_{WW}=- \Delta g^\prime_W
\end{align}
The equalities above hold exactly when, in the most right-hand side,
only the linear ${\cal O}(\Lambda^{-2})$ from dimension-six operators
and ${\cal O}(\Lambda^{-4})$ from dimension-eight operators are
included in the {\sl overline coefficients}.  In the same form the
correction to the $W$ mass obtained from Eq.~\eqref{eq:mwoblique}
coincides with Eq.~\eqref{eq:dmwt} when the dimension-six square terms
are not included. %\medskip

We finish by stressing that in the context of USMEFT, the
parametrization of the universal effects in terms of oblique
parameters obtained by linearly expanding the gauge boson
self-energies does not provide a consistent series in $(1/\Lambda)$
beyond $(1/\Lambda^2)$. The consistent expansion requires the
inclusion of the terms quadratic in the Wilson coefficients as
presented in Sec.~\ref{sec:formalism}.  Nevertheless, as expected, the
amplitudes obtained in terms of oblique parameters match the full expresions
for the terms linear in the operator coefficients and allows for a
clean identification of the number of independent combinations of
operator coefficients.

%%%%%%%%%%%%%%%%%%%%%%%%%%%%%%%%%%%%%%%%%%%%%%%%%%%%%%%%%%%%%%%%%%%%%%
\section{Analysis Framework}
\label{sec:framework}

In this work, our goal is to study the constraints on the USMEFT
Wilson coefficients imposed by neutral- and charged-current Drell-Yan
processes in combination with EWPO. %\medskip

Regarding the Drell-Yan processes, the larger LHC energy of
the 13 TeV runs implies that these runs are more sensitive to the
presence of anomalous couplings, as expected. Unfortunately only very
recently the ATLAS collaboration has presented a dedicated study of
Drell-Yan CC process at 13 TeV~\cite{ATLAS:2025hhn}. No dedicated
study of Drell-Yan NC process at this energy has been published with
detailed enough information on the differential cross sections to allow
for analysis outside of the collaborations. Determination of the
Drell-Yan NC cross sections have been presented only using 8 TeV
data~\cite{ATLAS:2016gic,CMS:2014jea}. However, both ATLAS and CMS
have searched for new resonances in the $\ell^+\ell^-$ and
$\ell^\pm \nu_\ell$ channels with the full 13 TeV luminosity. These
searches can be recast into bounds on the USMEFT by studying their data on the
lepton pair invariant mass distribution and the transverse mass
spectrum respectively. As such, we have included the high invariant mass
part of those distributions in the analyses, and for convenience, we
have also rebinned the data to guarantee a minimum number of events
per bin. We present in Table~\ref{tab:dydata} the summary of data we
include in the analyses. %\medskip

%%%%%%%%%%%%%%%%%%%%%%%%%%%%%%%%%%%%%%%%%%%%%%%%%%%%%%%
\begin{table}
\begin{tabular}{|c|c|c|c|c|c|c|}
  \hline
  Channel & Distribution & \# bins & ranges  & data set & Int. Lum.
  \\
  \hline
  NC  & $\frac{d^2\sigma}{dm_{\ell\ell}d|y|_{\ell\ell}}$ & 48&
  $116 \text{\;GeV} \leq m_{\ell\ell}\leq 1.5 \text{\;TeV}$
  $0 \leq y_{\ell\ell}\leq 2.4$ 
  &ATLAS 8 TeV  & 20.3 fb$^{-1}$~\cite{ATLAS:2016gic}
  \\
  \hline
  NC &  $\frac{dN_\text{ev}}{dm_{e^+e^-}}$ & 20 & $250 \text{\;GeV} \leq m_{e^+e^-}\leq 5 \text{\;TeV}$ &
  ATLAS 13 TeV & 139 fb$^{-1}$~\cite{ATLAS:2019erb}
\\
  NC &  $\frac{dN_\text{ev}}{dm_{\mu^+\mu^-}}$ & 20 & $250 \text{\;GeV} \leq m_{\mu^+\mu^-}\leq 5 \text{\;TeV}$ &
  ATLAS 13 TeV & 139 fb$^{-1}$~\cite{ATLAS:2019erb}
\\
NC &  $\frac{dN_\text{ev}}{dm_{e^+e^-}}$ & 20 & $300 \text{\;GeV} \leq m_{e^+e^-}\leq 6 \text{\;TeV}$ 
& CMS 13 TeV & 137 fb$^{-1}$~\cite{CMS:2021ctt}
\\
  NC &  $\frac{dN_\text{ev}}{dm_{\mu^+\mu^-}}$ & 20 & $300 \text{\;GeV} \leq m_{\mu^+\mu^-}\leq 7 \text{\;TeV}$ &
 CMS 13 TeV & 137 fb$^{-1}$~\cite{CMS:2021ctt}\\
  \hline
  CC & $\frac{d\sigma}{dm_T}$ & 20 & $200 \text{\;GeV} \leq m_{T,\ell \nu}
  \leq 5 \text{\;TeV}$ & ATLAS 13 TeV & 140 fb$^{-1}$~\cite{ATLAS:2025hhn}                                                      \\  
CC & $\frac{d N}{dm_T}$ & 20 & $440 \text{\;GeV} \leq m_{T,e
                               \nu}\leq 7 \text{\;TeV}$ & CMS 13 TeV &
                                                                      138 
                                                          fb$^{-1}$~\cite{CMS:2022krd} \\
CC & $\frac{d N}{dm_T}$ & 20 & $600 \text{\;GeV} \leq m_{T,\mu\nu}\leq 7
                               \text{\;TeV}$ & CMS 13 TeV & 138 
                                                          fb$^{-1}$~\cite{CMS:2022krd} \\
  \hline
\end{tabular} 
\caption{Neutral- and charged-current Drell-Yan data considered in our
  analyses.}
\label{tab:dydata}
\end{table}
%%%%%%%%%%%%%%%%%%%%%%%%%%%%%%%%%%%%%%%%%%%%%%%%%%%%%%%

As discussed in Sec.~\ref{sec:formalism} a total of six dimension-six
operators and eleven dimension-eight operators contribute to the
amplitudes at order $1/\Lambda^4$ of which we have identified eight
combinations of Wilson coefficients entering linearly, the five {\sl
  overline coefficients} $\overline\Delta_{4F}$, $\overline{c}_{BW}$,
$\overline{c}_{\Phi,1}$, $\overline{c}^{3}_{W^2 H^4}$, and
$\overline{c}_{2JB}$ in Eqs.~\eqref{eq:d4fb}--~\eqref{eq:c3whb}, and
Eq.~\eqref{eq:c2jbb}, together with the coefficients of the
dimension-eight operators $c^{(7)}_{\psi^4 H^2}$,
$c^{(2)}_{\psi^4 D^2}$ , and $c^{(3)}_{\psi^4 D^2}$.  In addition
there are contributions purely quadratic in the four dimension-six
coefficients $\Delta_{4F}$, ${c}_{BW}$, ${c}_{\Phi,1}$, and
${c}_{2JB}$.  %\medskip

The theoretical predictions needed for the analyses were obtained with
MadGraph5\_aMC@NLO~\cite{Frederix:2018nkq} at leading-order in QCD and
QED, with the UFO files for the effective Lagrangian with the seventeen
operators of the rotated basis generated with
FeynRules~\cite{Christensen:2008py,
  Alloul:2013bka} including also the ${\cal O}(\Lambda^{-2})$ and
${\cal O}(\Lambda^{-4})$ terms from the finite renormalization of the
SM inputs.  Parton shower and hadronization was performed using
PYTHIA8~\cite{Sjostrand:2007gs}, and the fast detector simulation was
carried out with Delphes~\cite{deFavereau:2013fsa}. Jet analyses was
done using FASTJET~\cite{Cacciari:2011ma}. Exclusively for the ATLAS NC
data~\cite{ATLAS:2019erb}, the detector response was simulated using
Rivet~\cite{Bierlich:2019rhm, Buckley:2019stt}, with the analysis code
provided by the experimental collaboration.  For the ATLAS 8 TeV data
\cite{ATLAS:2016gic}, QCD NNLO corrections for the SM predictions
  were incorporated using MATRIX~\cite{Grazzini:2017mhc}. When
  required, we corrected these predictions bin by bin by the SM
  correspondent k-factors for higher order QCD corrections.%\medskip

In order to address the dependence of the results on the order of the
expansion, we have performed the analyses at different orders. We
label the different analyses as

\begin{itemize}
  
\item ${\cal O}(\Lambda^{-4})$: Including contributions from
  dimension-six and dimension-eight operators and keeping their
  effects in the observables up to quadratic order in the
  dimension-six operator coefficients and linear order in the
  dimension-eight operator coefficients. This is the main focus of our
  work.  In addition we make the following analyses for comparison:

\item ${\cal O}(\Lambda^{-2})$: Including only contributions from
  dimension-six operators and keeping their effects in the observables
  at linear order in the operator coefficients -- that is, considering
  only the first two terms in Eq.~\eqref{eq:ampli}.

\item Dim-6 + (Dim-6)$^2$: Including only contributions from
  dimension-six operators and keeping their effects in the observables
  up to quadratic order in the operator coefficients, {\em i.e.}  not
  including the last term in Eq.~\eqref{eq:ampli}.
  
\item Dim-6 + Dim-8: Including contributions from dimension-six and
  dimension-eight operators and keeping their effects in the
  observable only at linear order in all operator coefficients.  This
  means including neither the third nor fourth term of
  Eq.~\eqref{eq:ampli}.  We present the results of this analysis also
  in terms of the generalized oblique parameters.

\end{itemize}

At ${\cal O}(\Lambda^{-4})$ we find that even including the quadratic
contributions, the analysis cannot break the degeneracies between the
different Wilson coefficients entering in the {\sl overline coefficients}.
Thus we proceed by substituting $\Delta_{4F}$, ${c}_{BW}$, ${c}_{\Phi,1}$, and
${c}_{2JB}$ with $\overline{\Delta}_{4F}$, $\overline{c}_{BW}$,
$\overline{c}_{\Phi,1}$, and $\overline{c}_{2JB}$ in the quadratic
terms and in the process neglect terms of ${\cal O}(\Lambda^{-6})$.
With this we define a $\chi^2$ function which depends on the
eight Wilson coefficient combinations
\begin{equation}
  \chi^2_\text{DY,NC}(
  \overline{c}_{BW} , \overline{c}_{\Phi,1},\overline{\Delta}_{4F},
  \overline{c}_{2JB}, c^{(2)}_{\psi^4 D^2}, c^{(3)}_{\psi^4 D^2}, c^{(7)}_{\psi^4 H^2})+
\chi^2_\text{DY,CC}(\overline{c}_{BW} ,
\overline{c}_{\Phi,1},\overline{\Delta}_{4F},\overline{c}^{(3)}
_{W^2H^4}, c^{(3)}_{\psi^4 D^2}) \;.
\end{equation}

With respect to the EWPO, we include 12 $Z$-pole
observables~\cite{ALEPH:2005ab}: $\Gamma_Z$, $\sigma_{h}^{0}$,
${\cal A}_{\ell}(\tau^{\rm pol})$, $R^0_\ell $,
${\cal A}_{\ell}({\rm SLD})$, $A_{\rm FB}^{0,l}$ $R^0_c$, $R^0_b$,
${\cal A}_{c}$,${\cal A}_{b}$, $A_{\rm FB}^{0,c}$, and
$A_{\rm FB}^{0,b}$ and two $W$ pole observables $M_W$ and $\Gamma_W$
taken from~\cite{ParticleDataGroup:2024cfk}. Notice that the average
leptonic $W$ branching ratio is not included because it does not lead
to any additional constraint on universal EFT's.  We include in our
analyses the correlations among these inputs, as given in
Ref.~\cite{ALEPH:2005ab}, and the SM predictions and their
uncertainties due to variations of the SM parameters were extracted
from~\cite{deBlas:2022hdk}.  %\medskip

As discussed in Sec.~\ref{sec:formalism} the $Z$ and $W$ pole
observables to order ${\cal O}(\Lambda^{-4})$ which are dominant can
be expressed in terms of four combinations $\widetilde{c}_{BW}$,
$\widetilde{c}_{\Phi,1}$, $\widetilde{c}^{(3)}_{W^2H^4}$, and
$\widetilde{\Delta}_{4F}$ which involve six of the eight combinations
testable in Drell-Yan: $\overline{c}_{BW}$, $\overline{c}_{\Phi,1}$,
$\overline{c}^{(3)}_{W^2H^4}$, $\overline{\Delta}_{4F}$,
$c^{(2)}_{\psi^4D^2}$, and $c^{(3)}_{\psi^4D^2}$.  As with the Drell-Yan
observables, the dimension-six square contributions to EWPO are not able to
break the degeneracies, thus we can proceed by  neglecting terms of
${\cal O}(\Lambda^{-6})$ and we write the couplings relevant to the
EWPO as Eqs.~\eqref{eq:dgZ}, ~\eqref{eq:dgW} where we substitute
$\Delta_{4F}$, $c_{BW}$, and $c_{\Phi,1}$ with $\widetilde{\Delta}_{4F}$
$\widetilde{c}_{BW}$ and $\widetilde{c}_{\Phi,1}$ in
$\Delta g_i^\square$ and $(\Delta M_W)^\square$.  With this, the EWPO
chi-squared function is:
\begin{eqnarray}
\chi^2_\text{EWPO}(\overline{c}_{BW} , \overline{c}_{\Phi,1},\overline{\Delta}_{4F},
\overline{c}^{(3)}_{W^2H^4},c^{(2)}_{\psi^4 D^2}, c^{(3)}_{\psi^4 D^2})&\equiv& \chi^2_{\rm EWPO}
  (\widetilde{c}_{BW},\widetilde{c}_{\Phi,1},
  \widetilde{c}^{(3)}_{W^2H^4}, \widetilde{\Delta}_{4F}) \; .
\label{eq:chiewpo2}
\end{eqnarray}
which can effectively only constrain the four {\sl tilde coefficients} in
Eq.~\eqref{eq:tildas}.

%%%%%%%%%%%%%%%%%%%%%%%%%%%%%%%%%%%%%%%%%%%%%%%%%%%%%%%%%%%%%%%%%%%%%%
\section{Results}
\label{sec:results}

\begin{figure}%\centering
 \includegraphics[width=0.65\textwidth]{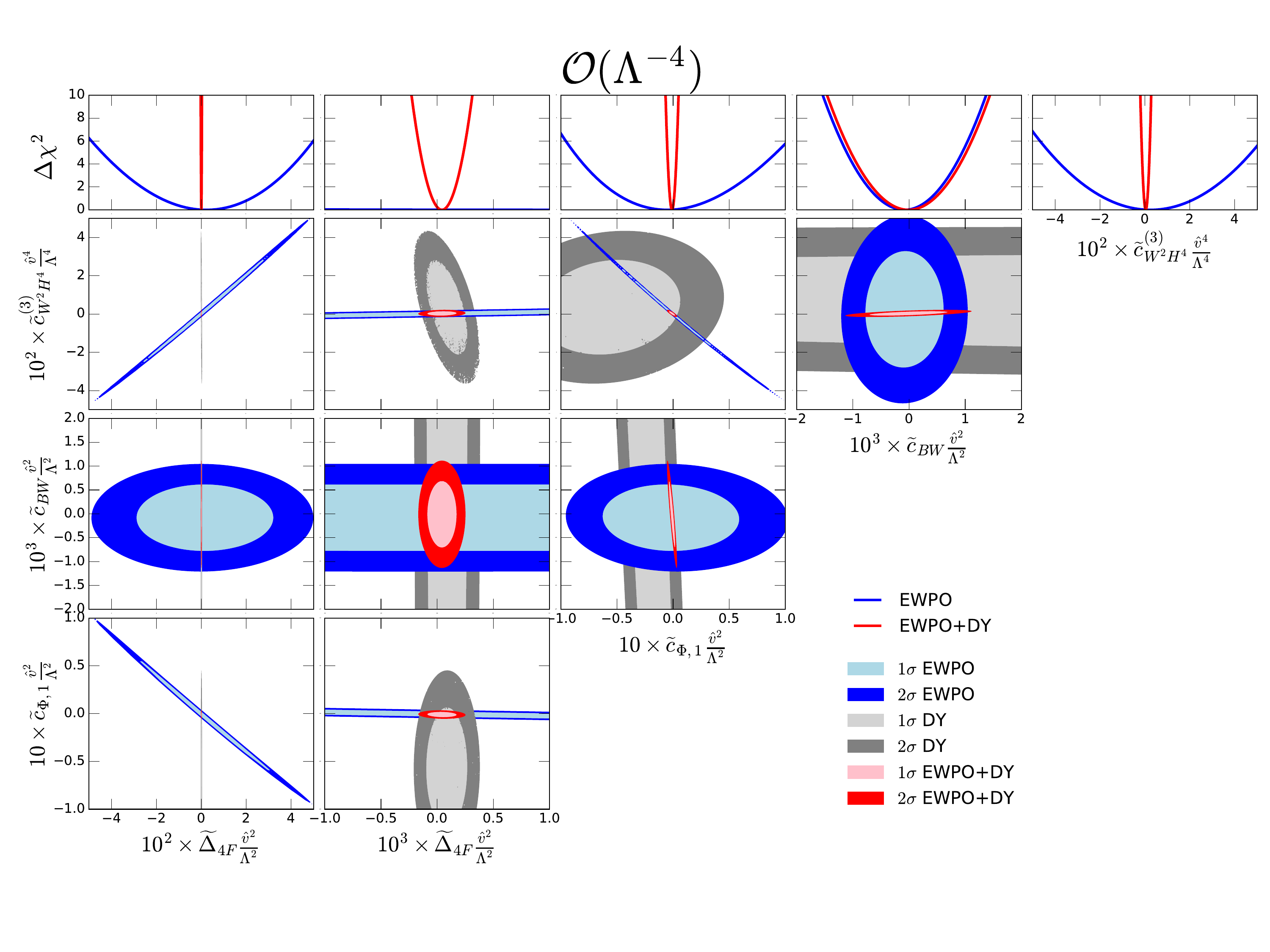}\hspace*{-1cm}
\includegraphics[width=0.4\textwidth]{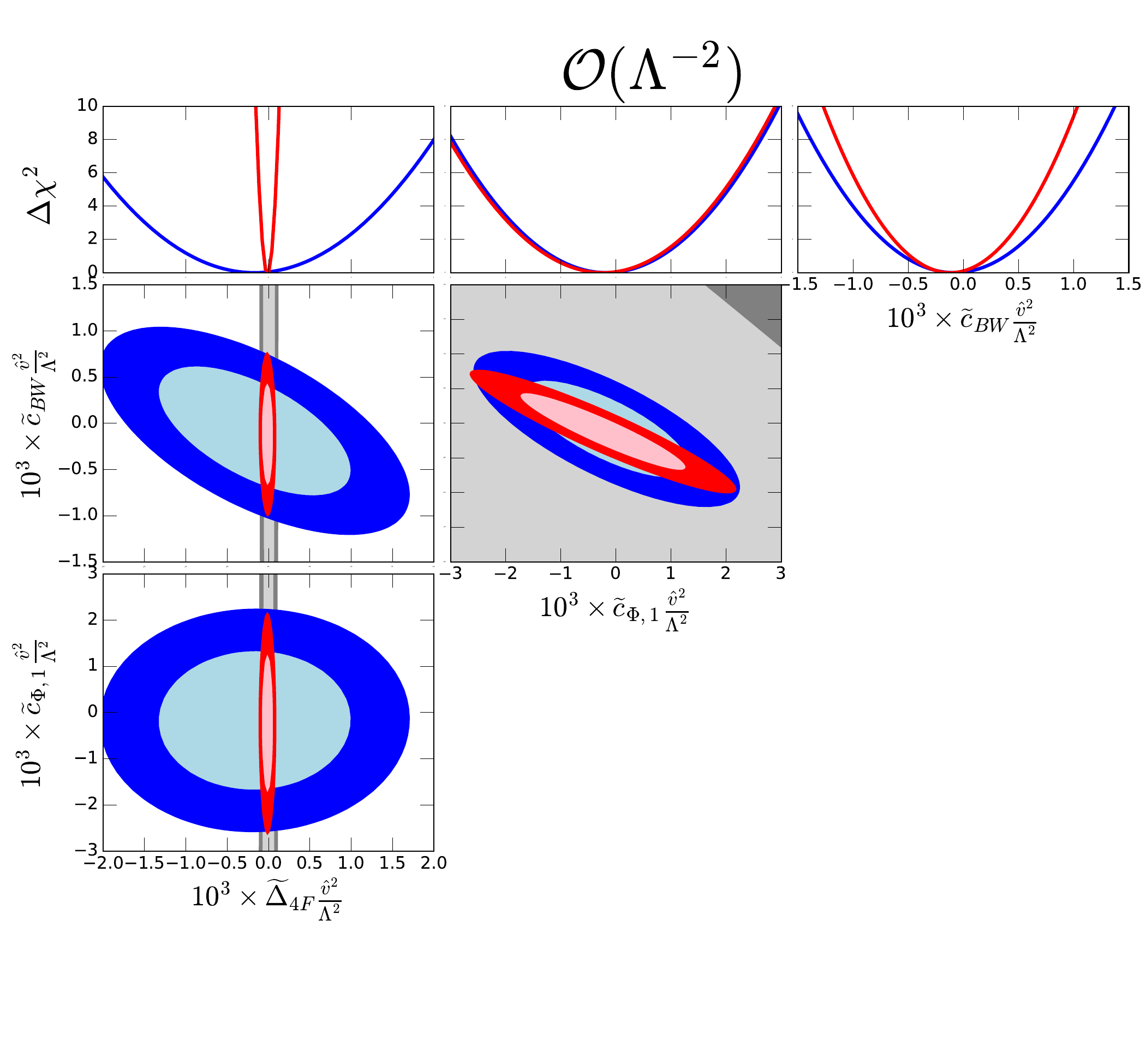}  
\caption {One- and two-dimensional projections of $\Delta{\chi}^2$
  from the analyses of EWPO and Drell-Yan data performed to
  ${\cal O} (\Lambda^{-4})$ on the left ( ${\cal O} (\Lambda^{-2})$ on
  the right) for the coefficients
  $\widetilde{c}_{BW} \vh^2/\Lambda^2$,
  $\widetilde{c}_{\Phi,1} \vh^2/\Lambda^2$,
  $\widetilde \Delta_{4F} \vh^2/\Lambda^2$, and
  $\widetilde c^{(3)}_{W^2H^4}\vh^4/\Lambda^4$
  ($\widetilde{c}_{BW} \vh^2/\Lambda^2$,
  $\widetilde{c}_{\Phi,1} \vh^2/\Lambda^2$, and
  $\widetilde \Delta_{4F} \vh^2/\Lambda^2$), as indicated in each
  panel after marginalizing over the 7/6 (2/1 at ${\cal O} (\Lambda^{-2})$) undisplayed
  parameters for one- and two-dimensional projections respectively. 
  Notice that the second column in the left figure is a
  magnification of the results on the first column for better
  visibility as well as the change of scale in the axis between the
  panels on the left and on the right figure.}
  \label{fig:ewpd8}
\end{figure}

We start by studying the complementarity and improvement on the
sensitivity between the DY results and the EWPO. In order to do so we
project the results of the analysis over the {\sl tilde
  coefficients}. The result is shown in the left in
Fig.~\ref{fig:ewpd8} which contains the one- and two-dimensional
projections of $\Delta\chi^2$ functions for the
${\cal O}(\Lambda^{-4})$ analyses of the EWPO and DY separately and
their combination as a function of the four tilde parameters.  The top
row contains the one-dimensional marginalized projections of
$\Delta\chi^2$'s as a function of the four combinations of Wilson
coefficients. From the analysis we obtain the 95\% CL allowed ranges for the
coupling combinations taking part in the EWPO listed in
Table~\ref{tab:limitsewpo}. %\medskip

The lower panels of Fig.~\ref{fig:ewpd8} depict the two-dimensional
$1\sigma$ and $2\sigma$ allowed regions for the different
analyses. For comparison we show the equivalent results for the
${\cal O}(\Lambda^{-2})$ analysis on the right which involves
only three coefficients.  First,  comparing the result from the analyses
of the EWPO (blue regions) in Fig.~\ref{fig:ewpd8} we observe the
wider allowed range of parameters $\widetilde{c}_{\Phi,1}$,
$\widetilde{\Delta}_{4F}$ in the ${\cal O}(\Lambda^{-4})$ analysis, and
the well-known strong correlations among $\widetilde{c}_{\Phi,1}$,
$\widetilde{\Delta}_{4F}$ and $\widetilde{c}^{(3)}_{W^2H^4}$. These
correlations are due to the cancellation of their linear contributions
to the $Z$ couplings and $W$ mass when
\begin{equation}
\widetilde{c}_{\Phi,1}=-2\widetilde{\Delta}_{4F}=-2
  \widetilde{c}^{(3)}_{W^2H^4} \frac{\vh^2}{\Lambda^2}\;.
\end{equation}
Along this direction the bounds on these three combinations dominantly
come from $\Gamma_W$ which is less precisely determined.  This
correlation weakens the limits on the Wilson
coefficient combinations shown in Table~\ref{tab:limitsewpo}  by a factor of $2-3$ with
respect to the order ${\cal O}(\Lambda^{-2})$ analysis. %\medskip

%%%%%%%%%%%%%%%%%%%%%%%%%%%%%%%%%%%%%%%%%%%%%%%%%%%%%%%
%\begin{figure}%\centering 
%  \caption {Same as Fig.~\ref{fig:ewpd8} but for analysis performed to
%   ${\cal O} (\Lambda^{-2})$ which involves 3 parameters.}
%  \label{fig:ewpd6}
%\end{figure}
%%%%%%%%%%%%%%%%%%%%%%%%%%%%%%%%%%%%%%%%%%%%%%%%%%%%%%%

%%%%%%%%%%%%%%%%%%%%%%%%%%%%%%%%%%%%%%%%%%%%%%%%%%%%%%%
\begin{table}
  \begin{tabular}{|c||c|c||c|c|}
\hline
    coupling & \multicolumn{4}{c|}{95\% CL allowed range}
               \\
               \hline
    & \multicolumn{2}{c|}{${\cal O}(\Lambda^{-2})$}
    & \multicolumn{2}{c|}{${\cal O}(\Lambda^{-4})$}\\\hline            
& EWPO & EWPO+DY & EWPO & EWPO+DY \\\hline
               $\frac{\vh^2}{\Lambda^2} \widetilde{c}_{BW}$
               & $[-10,8.4]\times 10^{-4}$
               & $[-8.5,6.1]\times 10^{-4}$
               & $[-10,8.4]\times 10^{-4}$
               & $[-9.3,9.1]\times 10^{-4}$
    \\
    $\frac{\vh^2}{\Lambda^2} \widetilde{c}_{\Phi,1}$
    & $[-2.1,1.8]\times 10^{-3}$
    & $[-2.2,1.7]\times 10^{-3}$
    & $[-8.0,8.1]\times 10^{-2}$
    & $[-4.6,2.7]\times 10^{-3}$
    \\ 
    $\frac{\vh^2}{\Lambda^2} \widetilde{\Delta}_{4F}$
    & $[-1.7,1.4]\times 10^{-3}$
    & $[-10,7.6]\times 10^{-5}$
    & $[-3.9,4.1]\times 10^{-2}$
    & $[-1.3,2.1]\times 10^{-4}$
    \\
    $\frac{\vh^4}{\Lambda^4} \widetilde{c}_{W^2H^4}^{(3)}$
    &  & 
    & $[-3.8,4.3]\times 10^{-2}$
    & $[-1.2,1.9]\times 10^{-3}$    
    \\
 \hline
  \end{tabular}
  \caption{95\% CL allowed ranges for the effective couplings entering the EWPO analysis.}
  \label{tab:limitsewpo}
\end{table}
%%%%%%%%%%%%%%%%%%%%%%%%%%%%%%%%%%%%%%%%%%%%%%%%%%%%%%%

In Fig.~\ref{fig:ewpd8} we also see that the analysis of Drell-Yan
data by itself provided a two orders of magnitude stronger constraint
on the coefficient $\widetilde{\Delta}_{4F}$ which contains the
four-fermion dimension-six operator coefficient $c_{2JW}$.  This
effect of {\sl energy helping accuracy} discussed in
Ref.~\cite{Farina:2016rws} in the context of an
${\cal O}(\Lambda^{-2})$ analysis arises from the contribution of the
operator $Q_{2JW}$ to the Drell-Yan four-fermion contact amplitudes;
see Eqs.~\eqref{eq:NGG}--~\eqref{eq:NWW}, or equivalently to the
modified propagators~\eqref{eq:eepgg}--~\eqref{eq:eepww} which
dominate at higher invariant masses.  Interestingly, in the
${\cal O}(\Lambda^{-4})$ analysis the resulting effect in the
combination of the DY with the EWPO is quantitatively more relevant
because DY results contribute to breaking the very strong degeneracy
present in the EWPO analysis. As a consequence, as seen in
Fig.~\ref{fig:ewpd8} and in Table~\ref{tab:limitsewpo} the inclusion
of DY results not only in a better determination of
$\widetilde{\Delta}_{4F}$, but also
constraints on $\widetilde{c}_{\Phi,1}$ and
$\widetilde{c}^{(3)}_{W^2H^4}$ of a factor $\sim 20$ stronger.  %\medskip

%%%%%%%%%%%%%%%%%%%%%%%%%%%%%%%%%%%%%%%%%%%%%%%%%%%%%%%
\begin{figure}%\centering
  \includegraphics[width=0.8\textwidth]{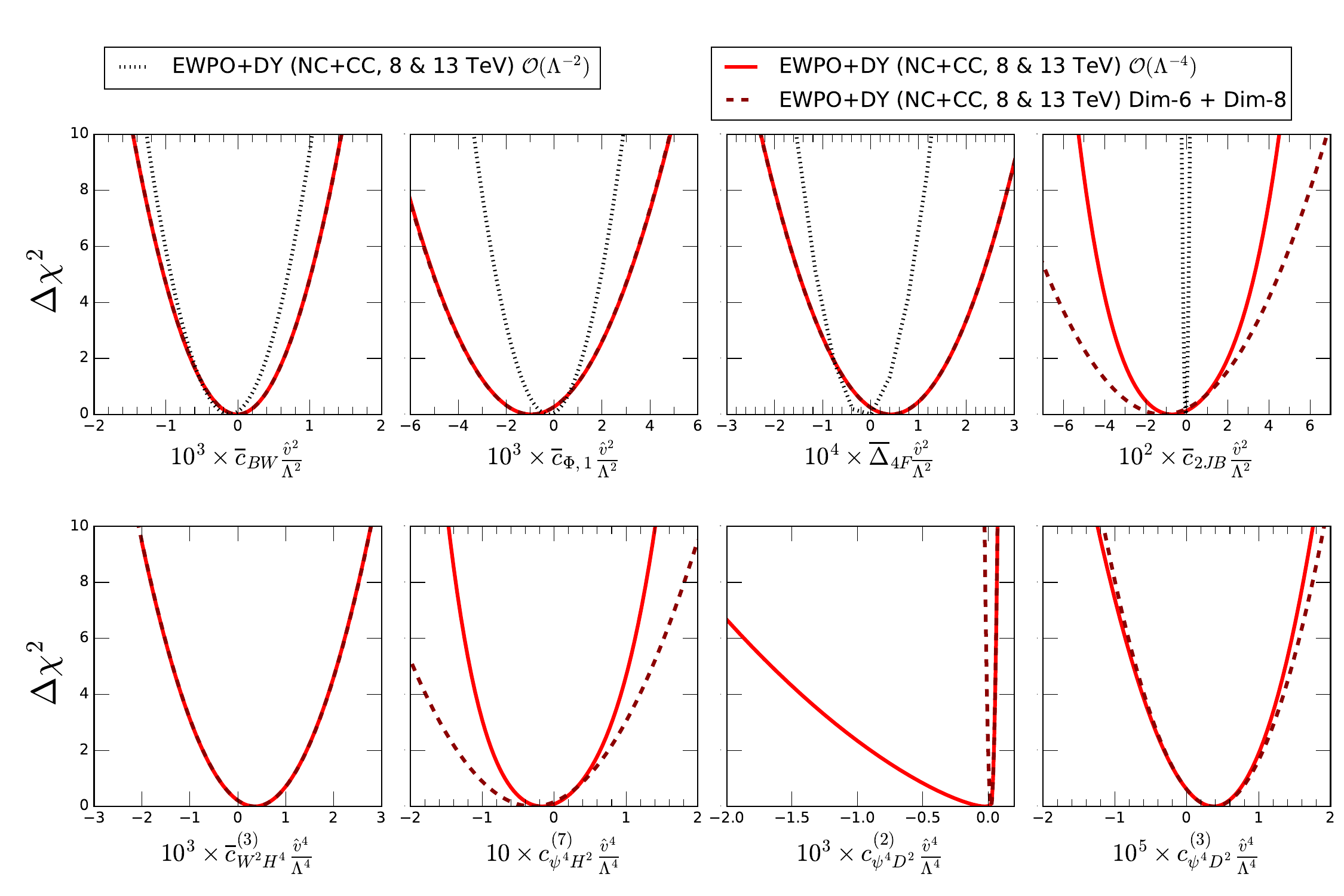}
  \caption {One-dimensional projections of $\Delta{\chi}^2$ after
    marginalization over all other coefficients for the different
    analyses performed as labeled in the figure.}
\label{fig:compa1d}
\end{figure}
%%%%%%%%%%%%%%%%%%%%%%%%%%%%%%%%%%%%%%%%%%%%%%%%%%%%%%%

%%%%%%%%%%%%%%%%%%%%%%%%%%%%%%%%%%%%%%%%%%%%%%%%%%%%%%%
\begin{figure}%\centering
  \includegraphics[width=0.9\textwidth]{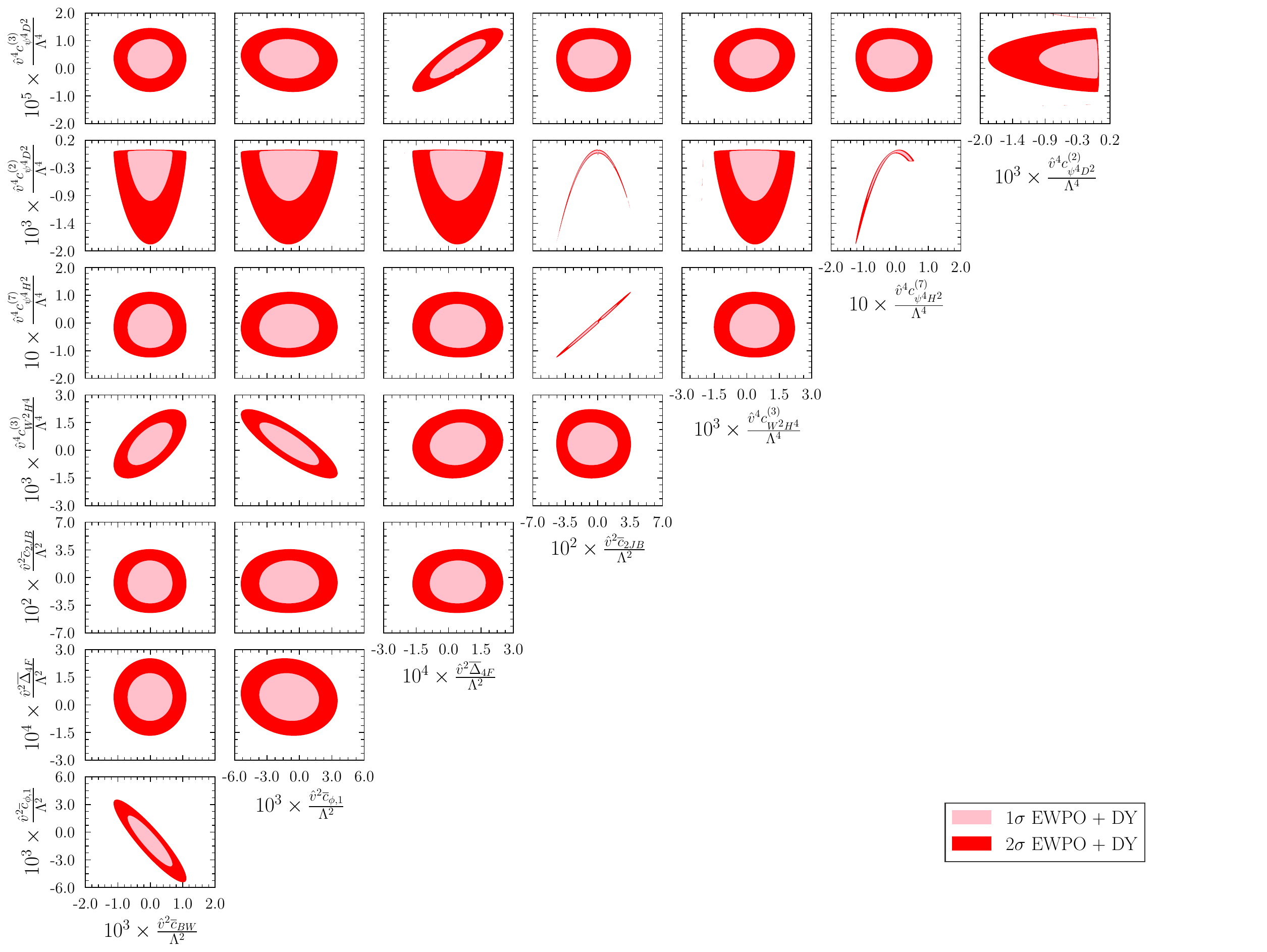}
\caption{Two-dimensional projection of the 
$\Delta{\chi}^2$ from the analysis
  of EWPO+Drell-Yan data for the ${\cal O}(\Lambda^{-4})$ analysis.}
\label{fig:contbar}
\end{figure}
%%%%%%%%%%%%%%%%%%%%%%%%%%%%%%%%%%%%%%%%%%%%%%%%%%%%%%%

%%%%%%%%%%%%%%%%%%%%%%%%%%%%%%%%%%%%%%%%%%%%%%%%%%%%%%%
\begin{table}
  \begin{tabular}{|c||c|c||c|c|}
\hline
    coupling & \multicolumn{4}{c|}{95\% CL allowed range EWPO+DY}
    \\\hline
    & ${\cal O}(\Lambda^{-2})$ &  Dim-6+(Dim-6)$^2$ 
    & ${\cal O}(\Lambda^{-4})$ &   Dim-6+Dim-8\\\hline
    $\overline{c}_{BW}\frac{\vh^2}{\Lambda^2}$
    &
 $[  -8.5 ,   6.1 ]\times 10^{-4 }$&
 $[  -8.4 ,   6.1 ]\times 10^{-4 }$&
    $[  -9.3 ,   9.1 ]\times 10^{-4 }$&
    $[  -9.3 ,   9.1]\times 10^{-4 }$\\
$\overline{c}_{\Phi,1}\frac{\vh^2}{\Lambda^2}$&    
 $[  -2.2 ,   1.7]\times 10^{-3}$&
 $[  -2.2 ,   1.7 ]\times 10^{-3 }$&
    $[  -4.6 ,   2.7 ]\times 10^{-3}$&
    $[  -4.6 ,   2.7 ]\times 10^{-3 }$\\
$\overline{\Delta}_{4F}\frac{\vh^2}{\Lambda^2}$    &    
 $[  -10 ,   7.6 ]\times 10^{-5 }$&
 $[  -10 ,   7.8 ]\times 10^{-5 }$&
    $[  -1.3 ,   2.1 ]\times 10^{-4 }$&
    $[  -1.3 ,   2.1 ]\times 10^{-4 }$\\
$\overline{c}_{2JW}\frac{\vh^2}{\Lambda^2}$    &    
 $[  -3.6 ,   4.8 ]\times 10^{-4 }$&
 $[  -3.7 ,   5.0 ]\times 10^{-4 }$&
    $[  -10 ,   6.1 ]\times 10^{-4 }$&
    $[  -10 ,   6.1 ]\times 10^{-4 }$\\
 $\overline{c}_{2JB}\frac{\vh^2}{\Lambda^2}$&       
 $[  -17 ,   7.9 ]\times 10^{-4 }$&
 $[  -21 ,   7.9 ]\times 10^{-4 }$&
    $[  -3.9 ,   2.9 ]\times 10^{-2 }$&
    $[  -6.2 ,   3.9 ]\times 10^{-2 }$\\
$\overline{c}^{(3)}_{W^2H^4}\frac{\vh^4}{\Lambda^4}$&        
   &  &
 $[  -1.4 ,   2.0 ]\times 10^{-3 }$&
    $[  -1.4 ,   2.0 ]\times 10^{-3 }$\\
 ${c}^{(7)}_{\psi^4H^2}\frac{\vh^4}{\Lambda^4}$&        
   &  &
 $[ -1.1    ,   0.93 ]\times 10^{-1 }$&
    $[ -1.8     ,  1.1     ]\times 10^{-1 }$\\
${c}^{(2)}_{\psi^4D^2}\frac{\vh^4}{\Lambda^4}$&        
   &  &
 $[  -14 ,   0.54 ]\times 10^{-4 }$&
    $[  -9.0 ,   54 ]\times 10^{-6 }$\\
${c}^{(3)}_{\psi^4D^2}\frac{\vh^4}{\Lambda^4}$&            
   &  &
 $[  -6.1,   13 ]\times 10^{-6 }$&
    $[  -5.9 ,   13]\times 10^{-6 }$\\\hline
  \end{tabular}
  \caption{95\%CL allowed ranges for the Wilson coefficients from the
    combined analysis of EWPO and DY for the different analysis
    assumptions. In all cases the ranges are obtained after
    marginalization over all other coefficients entering the analysis.
    For convenience we list in the third and fourth lines the bounds
    on the contributions from the $Q_{2JW}$ operator in terms of both
    $\overline {c}_{2JW}$ and
    $\overline{\Delta}_{4F} = -\frac{\eh^2}{2\sh^2} \overline
    {c}_{2JW}$.}
  \label{tab:rangesbar}
\end{table}
%%%%%%%%%%%%%%%%%%%%%%%%%%%%%%%%%%%%%%%%%%%%%%%%%%%%%%%

%%%%%%%%%%%%%%%%%%%%%%%%%%%%%%%%%%%%%%%%%%%%%%%%%%%%%%%
\begin{figure}%\centering
  \includegraphics[width=0.7\textwidth]{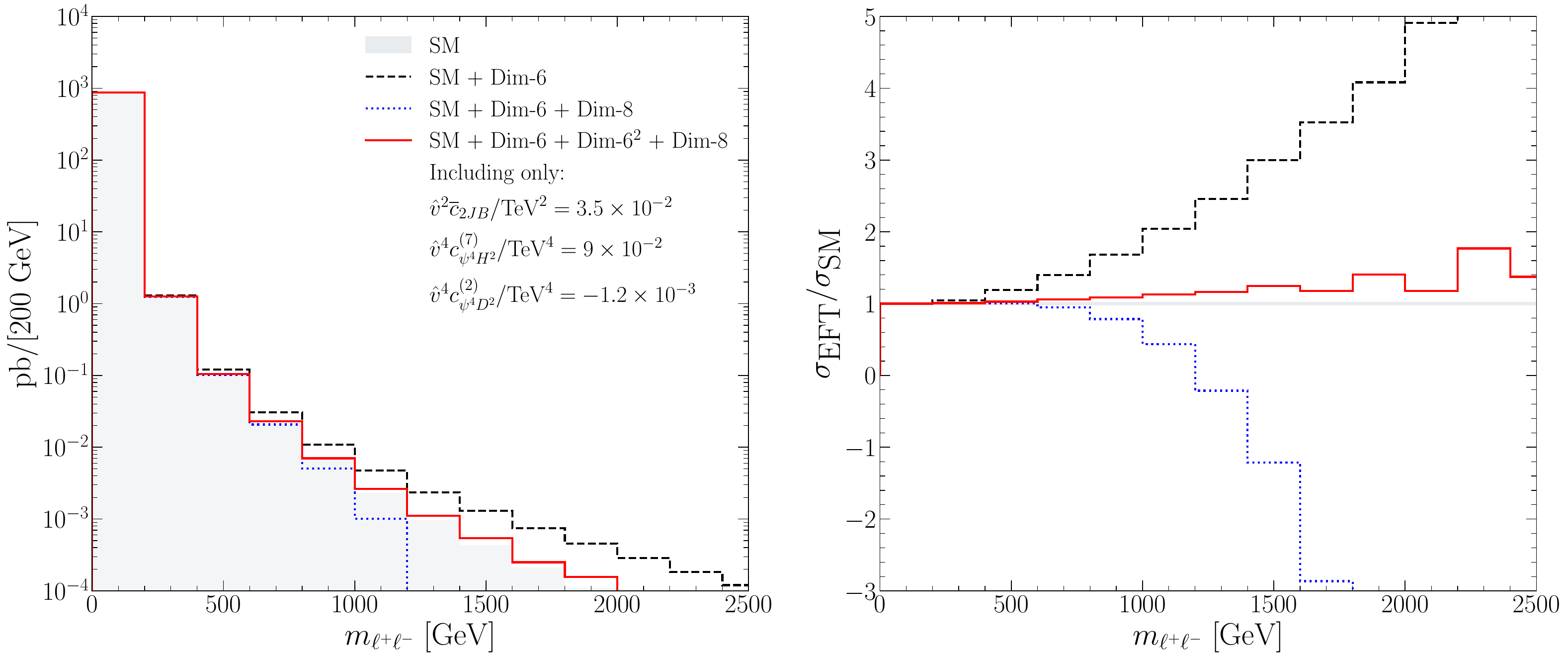}
  \caption{Prediction of the DY NC invariant mass distribution for
    several values of the Wilson coefficients as labeled in the
    figure.}
\label{fig:ncdist}
\end{figure}
%%%%%%%%%%%%%%%%%%%%%%%%%%%%%%%%%%%%%%%%%%%%%%%%%%%%%%%

%aqui

The results of the full analyses in terms of the eight {\sl overline
  coefficients} introduced in Sec.~\ref{sec:framework} are shown in
Figs.~\ref{fig:compa1d}
and~\ref{fig:contbar} and summarized in
Table~\ref{tab:rangesbar}. First, from Fig.~\ref{fig:compa1d} we see
that for none of the analyses presented do we find any favored deviation
from the SM predictions, and  the zero value for all of the parameters
lies at $\Delta\chi^2<1$. With respect to the allowed ranges,
comparing the results in the two
 left-most columns in Table~\ref{tab:rangesbar} we see that for
the analysis performed including only dimension-six operators, the
constraints on the coefficients $\overline{c}_{BW}$,
$\overline{c}_{\Phi,1}$, $\overline{c}_{2JW}$ (or equivalently
$\overline{\Delta}_{4F}$), and $\overline{c}_{2JB}$ are robust under
the inclusion of the dimension-six square contributions to the
observables.
Furthermore comparing these results with those of the analysis
performed including the dimension-eight operators (see two right
columns in Table~\ref{tab:rangesbar} and red curves in
Fig.~\ref{fig:compa1d}) we learn that the bounds on
$\overline{c}_{BW}$, $\overline{c}_{\Phi,1}$, $\overline{c}_{2JW}$,
($\overline{\Delta}_{4F}$) become slightly weaker while the constraint
on $\overline{c}_{2JB}$ becomes much weaker. This is a consequence of 
cancellations between dimension-six and dimension-eight contributions
which results into corelations between the allowed ranges. 

The correlations  in the determination of the coefficients is
expliclty shown  in Fig.~\ref{fig:contbar} where we plot two-dimensional
projections of the $\chi^2$ function after marginalizaiton over the other
six coefficients.  In the figure we see 
correlations between $\overline{c}_{BW}$ and $\overline{c}_{\Phi,1}$,
between $\overline{c}_{\Phi,1}$ and $\overline{c}^{3}_{W^2H^4}$ and
between $\overline{c}_{BW}$ and $\overline{c}^{3}_{W^2H^4}$. They are
related to those already observed among the corresponding {\sl
  tilde coefficients} in the ${\cal O}(\Lambda^{-4})$ results on the
left of Fig.~\ref{fig:ewpd8}.  We also observe a moderate correlation between
$\overline{\Delta}_{4F}$ ($\overline{c}_{2JW}$) and
$c^{(3)}_{\psi^4D^2}$.  We can trace its origin to the fact that
linear combinations of these two coefficients enter in both the pole
observables (see Eqs.~\eqref{eq:dg1t}--~\eqref{eq:dgwt}) and in the
four-fermion contact DY amplitudes in Eqs.~\eqref{eq:4FCC} and
~\eqref{eq:4FNC} (see also Eqs.\eqref{eq:NGG}--\eqref{eq:NWW} and
Eqs.~\eqref{eq:NPGG}--\eqref{eq:NPWW}).  However the relative effects
are different at the $Z$ and $W$ pole than in the DY amplitudes
because of the momentum dependence of the $c^{(3)}_{\psi^4D^2}$
contribution. Consequently the combination of EWPO and DY from both NC
and CC processes can independently bound the two coefficients
leaving only the correlation shown. %\medskip

From Fig.~\ref{fig:contbar} we also see that the most correlated bounds
correspond to the coefficients $\overline{c}_{2JB}$,
$c^{(7)}_{\psi^4 H^2}$, and $c^{(2)}_{\psi^4 D^2}$. We first observe a
very strong positive correlation between $\overline{c}_{2JB}$ and
$c^{(7)}_{\psi^4 H^2}$. These two operators do not contribute to EWPO
and only enter DY in the four-fermion NC contact amplitudes in
Eq.~\eqref{eq:4FNC} (see Eqs.\eqref{eq:NGG}--\eqref{eq:NGZ}), of which
$\overline{\cal N}_{\gamma\gamma}$ is numerically larger.
Consequently, the analysis provides the weakest bounds when
$\overline{\cal N}_{\gamma\gamma}$ cancels which occurs for
\begin{equation}
  \overline{c}_{2JB}\frac{\vh^2}{\Lambda^2}=\frac{\sh}{2\ch} c^{(7)}_{\psi^4 H^2} \frac{\vh^4}{\Lambda^4}
  \simeq 0.3 \,c^{(7)}_{\psi^4 H^2}\frac{\vh^4}{\Lambda^4} \;,
\label{eq:canceljb}  
\end{equation}
leading to the very strong correlation observed and the substantial
weakening of the bounds on $c_{2JB}$ when compared to the analysis
performed at ${\cal O}(\Lambda^{-2})$.
In addition we find very highly correlated non-elliptical allowed
regions for $\overline{c}_{2JB}$ and $c^{(2)}_{\psi^4 D^2}$ and also
for $c^{(7)}_{\psi^4 H^2}$ and $c^{(2)}_{\psi^4 D^2}$.  We trace this
behaviour to possible cancellations in the Drell-Yan NC distributions
between the linear contribution from negative values of
$c^{(2)}_{\psi^4 D^2}$ and the quadratic contribution from
$\overline{c}_{2JB}$ because both enter at ${\cal
  O}(\Lambda^{-4})$. We illustrate this behaviour in
Fig.~\ref{fig:ncdist} where we show the predicted invariant mass
distribution for DY NC cross section at 13 TeV for a set of values of
$\overline{c}_{2JB}$, $c^{(7)}_{\psi^4 H^2}$ and
$c^{(2)}_{\psi^4 D^2}$ within the 2$\sigma$ bounds from the
${\cal O}(\Lambda^{-4})$ analysis. As seen in the figure the full
${\cal O} (\Lambda^{-4})$ prediction (red line) is very similar to the
SM in the range of invariant masses shown. Conversely once the
dimension-six square contribution is not included (dotted blue line)
the prediction departs substantially from the SM.  Consequently,
bounds on negative values of $c^{(2)}_{\psi^4 D^2}$ are much stronger
if the dimension-six square contribution is not included as seen in
the dashed lines in Fig.~\ref{fig:compa1d} and the last column in
Table~\ref{tab:rangesbar}. This degeneracy in the DY event rates can
only be broken with larger statistics at the highest invariant
masses. %\medskip

%%%%%%%%%%%%%%%%%%%%%%%%%%%%%%%%%%%%%%%%%%%%%%%%%%%%%%%
\begin{figure}%\centering
  \includegraphics[width=0.9\textwidth]{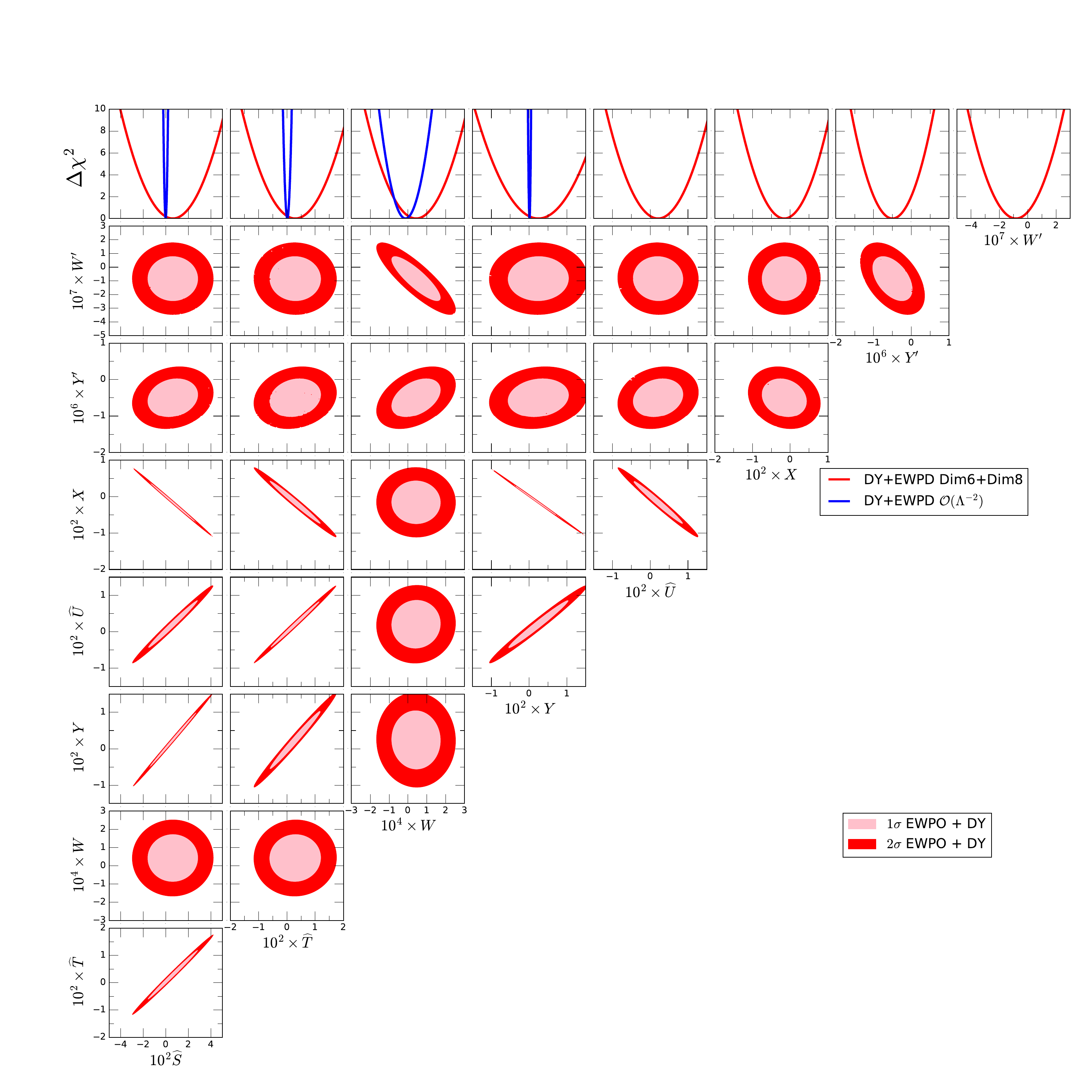}
  \caption{One- and two-dimensional projection of the $\Delta{\chi}^2$
    from the analysis of EWPO and Drell-Yan data in terms of the
    generalized oblique parameters.}
\label{fig:oblique}
\end{figure}
%%%%%%%%%%%%%%%%%%%%%%%%%%%%%%%%%%%%%%%%%%%%%%%%%%%%%%%

We finish by presenting in Fig.~\ref{fig:oblique} the results of an
analysis performed in terms of the generalized oblique parameters
introduced in Sec.~\ref{sec:oblique}. As discussed, in the framework
of the operator expansion this corresponds to an analysis which
neglects the dimension-six square effects. Therefore, by construction
the $\chi^2$ statistics are quadratic functions of all the oblique
parameters and the two-dimensional projections, are elliptic
regions. In this figure we see strong correlations among several of
the allowed ranges of the oblique parameters.  The correlations among
$\widehat{S}$, $\widehat{T}$ and $\widehat{U}$ dominantly stem from
the same effect as the correlation among $\overline{c}_{BW}$,
$\overline{c}_{\Phi,1}$ and $\overline{c}^{3}_{W^2H^4}$ observed in
Fig.~\ref{fig:contbar}, or equivalently among $\widetilde{c}_{BW}$,
$\widetilde{c}_{\Phi,1}$ and $\widetilde{c}^{3}_{W^2H^4}$ in
Fig.~\ref{fig:ewpd8}.  The anticorrelation between $W$ and $W'$ arises
from the same effects as $\overline{\Delta}_{4F}$
($\overline{c}_{2JW}$) and $c^{(3)}_{\psi^4D^2}$ discussed above.  In
addition we see a somewhat weak anticorrelation between $W'$ and
$Y'$ stemming from their dominant contribution to the tail of the
invariant mass distributions induced by $\epsilon'_{\gamma\gamma}$,
$\epsilon'_{\gamma Z}$, $\epsilon'_{ZZ}$, and $\epsilon'_{WW}$,
\eqref{eq:eepgg}--~\eqref{eq:eepww}. The correlations observed
involving $X$ and $Y$ mostly result from the cancellations equivalent to
that in Eq.~\eqref{eq:canceljb}; see Appendix~\ref{app:oblique} for
the expressions of the oblique parameters in terms of the operator
coefficients.  For the sake of comparison we also show in the figure
the one-dimensional projection of an analysis performed in terms of
the oblique parameters which are generated by USMEFT dimension-six
operators, $\widehat{S}$, $\widehat{T}$, $W$, and $Y$. From this
figure we can see that as a consequence of the above correlations, the
bounds on these four oblique parameters relax considerably when
including the effects of $\widehat{U}$, $X$, $W'$ and $Y'$ as
quantified in Table~\ref{tab:oblique}.%\medskip

%%%%%%%%%%%%%%%%%%%%%%%%%%%%%%%%%%%%%%%%%%%%%%%%%%%%%%%
%%%%%%%%%%%%%%%%%%%%%%%%%%%%%%%%%%%%%%%%%%%%%%%%%%%%%%%
\begin{table}
  \begin{tabular}{|c||c|c|}
\hline
  Parameter & \multicolumn{2}{c|}{95\% CL allowed range EWPO+DY}
\\\hline
& ${\cal O}(\Lambda^{-2})$ &  Dim-6+Dim-8\\\hline
${\widehat S}$
& $[  -1.5 ,   1.2 ]\times 10^{-3 }$&
  $[  -2.3 ,   3.6 ]\times 10^{-2 }$\\
${\widehat T}$  
 & $[  -8.4,   11 ]\times 10^{- 4}$&
  $[  -9.0 ,   15 ]\times 10^{- 3}$\\
$W$  
 & $[  -10,   7.6 ]\times 10^{-5 }$&
  $[  -1.3 ,   2.1 ]\times 10^{-4 }$\\
$Y$  
 & $[  -1.7 ,   3.7 ]\times 10^{-4}$&
  $[  -8.2 ,   13 ]\times 10^{-3 }$\\
$\widehat{U}$ & &
  $[  -7.7 ,   12]\times 10^{-3 }$\\
$X$ & &
  $[  -9.4 ,  6.3 ]\times 10^{- 3}$\\
$Y'$ &  &
  $[  -12 ,   2.0 ]\times 10^{-7 }$\\
 $W'$ & & $[  -3.0 ,   1.3 ]\times 10^{-7}$\\\hline
\end{tabular}
\caption{95\%CL allowed ranges for the oblique parameters from the
  combined analysis of EWPO and DY for the different analysis
  assumptions.  In all cases the ranges are obtained after
  marginalization over all other coefficients entering the analysis.}
\label{tab:oblique}  
\end{table}

%%%%%%%%%%%%%%%%%%%%%%%%%%%%%%%%%%%%%%%%%%%%%%%%%%%%%%%%%%%%%%%%%%%%%%
\section{Summary}
\label{sec:summary}

We have studied Drell-Yan production in universal theories
consistently including effects beyond those of USMEFT at
dimension-six.  We have focused on effects which are $C$ and $P$
conserving and found that eleven dimension-eight operators and six
dimension-six operators contribute to our analyses.  The chosen bases
are listed Tables~\ref{tab:uniopd6} and ~\ref{tab:uniopd8}.  Working
in the rotated basis in which operators with higher derivatives of the
bosonic fields have been replaced by the equations of motion in favor of
combinations of operators involving SM fermionic currents, we have
identified the eight combinations of the 17 Wilson coefficients which
are physically distinguishable by studying the invariant mass
distribution of the lepton pairs produced: $\overline\Delta_{4F}$,
$\overline{c}_{BW}$, $\overline{c}_{\Phi,1}$, $\overline{c}_{2JB}$,
and $\overline{c}^{3}_{W^2 H^4}$, given in
Eqs.~\eqref{eq:d4fb}--~\eqref{eq:c3whb} and Eq.~\eqref{eq:c2jbb},
together with the coefficients of the dimension-eight operators
$c^{(7)}_{\psi^4 H^2}$, $c^{(2)}_{\psi^4 D^2}$, and
$c^{(3)}_{\psi^4 D^2}$.  Of those eight, the four {\sl tilde
  coefficients} in Eq.~\eqref{eq:tildas} contribute EWPO at the $Z$
and $W$ poles. %\medskip

In Sec.~\ref{sec:oblique} we have introduced an extension of the
parametrization of universal effects in terms of 11 oblique parameters
obtained by linearly expanding the self-energies of the electroweak
gauge bosons to ${\cal O}(q^6)$. Of those, eight are generated by the
USMEFT at dimension-eight: $\Shat$, $\That$, $W$, $Y$, $\Uhat$, $X$,
plus two additional which we label $W'$ and $Y'$. The correspondence
between these eight oblique parameters and the operator coefficients
is given in Appendix~\ref{app:oblique}.%\medskip

We have performed combined analyses to a variety of LHC dilepton data
and the EWPO in order to quantify the constraints on the full
parameter space and studied the dependence of the derived constraints
with the order of the expansion considered.  We first have quantified
how the DY results can complement the constraints from EWPO on the
four {\sl tilde coefficients}. We found that in the
${\cal O}(\Lambda^{-4})$ analysis the resulting effect of the
combination of the DY with the EWPO is quantitatively more relevant
than at ${\cal O}(\Lambda^{-2})$ as, besides constraining 
the coefficient $\widetilde \Delta_{4F}$ better from contact four-fermion DY
amplitudes, DY results further contribute by breaking the very strong
degeneracy present in the EWPO analysis when including the
dimension-eight operators.  As a consequence, as seen in
Fig.~\ref{fig:ewpd8} and in Table~\ref{tab:limitsewpo} the inclusion
of the DY results results not only in the better determination of
$\widetilde{\Delta}_{4F}$ but also in a factor $\sim 20$ stronger
constraint on $\widetilde{c}_{\Phi,1}$, and
$\widetilde{c}^{(3)}_{W^2H^4}$.%\medskip

The results on the full eight parameter space are shown in
Figs.~\ref{fig:compa1d} and~\ref{fig:contbar} and
Table~\ref{tab:rangesbar}. They show that, when consistently including
all effects to ${\cal O}(\Lambda^{-4})$, the combination of EWPO and
DY provides robust constraints on the three coefficients with leading
dimension-six contributions $\overline{c}_{BW}$, $\overline{c}_{\Phi,1}$,
$\overline{c}_{2JW}$ which are only weaker by at most a factor $\sim$
2 with respect to the ${\cal O}(\Lambda^{-2})$ bounds.  Robust bounds
are also obtained for the dimension-eight operator coefficients
$\overline{c}^{(3)}_{W^2H^2}$ and $c^{(3)}_{\psi^4 D^2}$ which are not
affected by possible cancellations with dimension-six square
contributions. Conversely,  the bounds on the leading dimension-six
coefficient $\overline{c}_{2JB}$ is weakened by more than one order of
magnitude with respect to the ${\cal O}(\Lambda^{-2})$ limits due to
cancellations with the ${\cal O}(\Lambda^{-4})$ contributions from
$c^{(7)}_{\psi^4 H^2}$ and $c^{(2)}_{\psi^4 D^2}$. Consequently, the
bounds on these two dimension-eight Wilson coefficients, in particular
$c^{(2)}_{\psi^4 D^2}$ are the least robust.
Finally we have quantified the constraints on the eight oblique parameters
in Fig.~\ref{fig:oblique} and Table~\ref{tab:oblique}.

%%%%%%%%%%%%%%%%%%%%%%%%%%%%%%%%%%%%%%%%%%%%%%%%%%%%%%%

%%%%%%%%%%%%%%%%%%%%%%%%%%%%%%%%%%%%%%%%%%%%%%%%%%%%%%%
\acknowledgments

OJPE is partially supported by CNPq grant number 305762/2019-2 and
FAPESP grant 2019/04837-9.  M.M. is supported by FAPESP grant
2022/11293-8 and 2024/04246-9 while P.R. acknowledges support by FAPESP grant
2020/10004-7.  This project is funded by USA-NSF grant PHY-2210533.  It
has also received support from the European Union's Horizon 2020
research and innovation program under the Marie Sk\l odowska-Curie
grant agreement No 860881-HIDDeN, and Horizon Europe research and
innovation programme under the Marie Sk\l odowska-Curie Staff Exchange
grant agreement No 101086085 -- ASYMMETRY''.  It also receives support
from grants PID2019-105614GB-C21, and ``Unit of Excellence Maria de
Maeztu 2020-2023'' award to the ICC-UB CEX2019-000918-M, funded by
MCIN/AEI/10.13039/501100011033, and from grant 2021-SGR-249
(Generalitat de Catalunya).

%%%%%%%%%%%%%%%%%%%%%%%%%%%%%%%%%%%%%%%%%%%%%%%%%%%%%%%%%%%%%%%%%%%%%%
\bibliography{references}

\newpage
\appendix

%%%%%%%%%%%%%%%%%%%%%%%%%%%%%%%%%%%%%%%%%%%%%%%%%%%%%%%

%%%%%%%%%%%%%%%%%%%%%%%%%%%%%%%%%%%%%%%%%%%%%%%%%%%%%%%
\section{Universal bosonic operator basis at dimension-eight}
\label{app:boson8}
Murphy's basis contains 89 bosonic operators in total.  We list them in 
Table~\ref{tab:89murphy} together with their transformation properties
under $C$ and $P$ . In addition there are  86 bosonic operators
present in universal models which contain higher derivatives and
which can be rotated into fermionic operators by the SM EOM.
We list them in Table~\ref{tab:redundant} together
with their transformation properties
under $C$ and $P$. For convenience we indicate the 38+50 
operators that violate either $C$ or $P$ with  a lighter shade. %\medskip

%%%%%%%%%%%%%%%%%%%%%%%%%%%%%%%%%%%%%%%%%%%%%%%%%%%%%%%
\begin{table}
\scalebox{0.85}{
\begin{tabular}[t]{|l|l|c|c|c||l|l|c|c|c|}
  \hline    
  \multicolumn{2}{|c|}{\boldmath $ 1:X^4,\, X^3 X^{\prime}$} & C & P & CP&
  \multicolumn{2}{c|}{\boldmath  $1:X^2 X^{\prime 2}$}& C & P & CP
\\
\hline
$Q_{G^4}^{(1)}$ & $ (G_{\mu\nu}^A G^{A\mu\nu}) (G_{\rho\sigma}^B G^{B\rho\sigma})$ 
& \chm &\chm &\chm &
$Q_{G^2W^2}^{(1)} $ & $ (W_{\mu\nu}^I W^{I\mu\nu}) (G_{\rho\sigma}^A G^{A\rho\sigma})$
&\chm &\chm &\chm 
  \\
%%%
$Q_{G^4}^{(2)}$ & $ (G_{\mu\nu}^A \widetilde{G}^{A\mu\nu}) (G_{\rho\sigma}^B \widetilde{G}^{B\rho\sigma})$ 
& \chm &\chm &\chm &
  $Q_{G^2W^2}^{(2)}$ & $ (W_{\mu\nu}^I \widetilde{W}^{I\mu\nu}) (G_{\rho\sigma}^A \widetilde{G}^{A\rho\sigma})$
&\chm &\chm &\chm 
 \\
$Q_{G^4}^{(3)} $ & $ (G_{\mu\nu}^A G^{B\mu\nu}) (G_{\rho\sigma}^A G^{B\rho\sigma})$
  &\chm &\chm &\chm  &  
  $Q_{G^2W^2}^{(3)} $ & $ (W_{\mu\nu}^I G^{A\mu\nu}) (W_{\rho\sigma}^I G^{A\rho\sigma})$
  &\chm &\chm &\chm 
  \\
$Q_{G^4}^{(4)}$ & $ (G_{\mu\nu}^A \widetilde{G}^{B\mu\nu}) (G_{\rho\sigma}^A \widetilde{G}^{B\rho\sigma})$  
  &\chm &\chm &\chm & 
  $Q_{G^2W^2}^{(4)} $ & $ (W_{\mu\nu}^I \widetilde{G}^{A\mu\nu}) (W_{\rho\sigma}^I \widetilde{G}^{A\rho\sigma})$
  &\chm &\chm &\chm 
  \\
\cyan $Q_{G^4}^{(5)} $ & \cyan $ (G_{\mu\nu}^A G^{A\mu\nu}) (G_{\rho\sigma}^B \widetilde{G}^{B\rho\sigma})$
  &\chm  &\xm &\xm & 
  \cyan $Q_{G^2W^2}^{(5)} $ & \cyan $ (W_{\mu\nu}^I \widetilde{W}^{I\mu\nu}) (G_{\rho\sigma}^A G^{A\rho\sigma})$
  &\chm &\xm &\xm  
  \\
\cyan $Q_{G^4}^{(6)} $ & \cyan $ (G_{\mu\nu}^A G^{B\mu\nu}) (G_{\rho\sigma}^A \widetilde{G}^{B\rho\sigma})$
  &\chm & \xm & \xm &  
  \cyan $Q_{G^2W^2}^{(6)} $ & \cyan $ (W_{\mu\nu}^I W^{I\mu\nu}) (G_{\rho\sigma}^A \widetilde{G}^{A\rho\sigma})$
    &\chm &\xm &\xm  
    \\
$Q_{G^4}^{(7)} $ & $ d^{ABE} d^{CDE} (G_{\mu\nu}^A G^{B\mu\nu}) (G_{\rho\sigma}^C G^{D\rho\sigma})$
    &\chm &\chm &\chm & 
    \cyan $Q_{G^2W^2}^{(7)} $ & \cyan $ (W_{\mu\nu}^I G^{A\mu\nu}) (W_{\rho\sigma}^I \widetilde{G}^{A\rho\sigma})$
&\chm &\xm & \xm
\\
$Q_{G^4}^{(8)} $ & $ d^{ABE} d^{CDE} (G_{\mu\nu}^A \widetilde{G}^{B\mu\nu}) (G_{\rho\sigma}^C \widetilde{G}^{D\rho\sigma})$ 
  &\chm & \chm &\chm & 
  $Q_{G^2B^2}^{(1)} $ & $ (B_{\mu\nu} B^{\mu\nu}) (G_{\rho\sigma}^A G^{A\rho\sigma})$
  &\chm &\chm &\chm
\\
\cyan $Q_{G^4}^{(9)} $ & \cyan $ d^{ABE} d^{CDE} (G_{\mu\nu}^A G^{B\mu\nu}) (G_{\rho\sigma}^C \widetilde{G}^{D\rho\sigma})$ 
&\chm &\xm & \xm & 
$Q_{G^2B^2}^{(2)} $ & $ (B_{\mu\nu} \widetilde{B}^{\mu\nu}) (G_{\rho\sigma}^A \widetilde{G}^{A\rho\sigma})$
&\chm &\chm &\chm
\\
$Q_{W^4}^{(1)} $ & $ (W_{\mu\nu}^I W^{I\mu\nu}) (W_{\rho\sigma}^J W^{J\rho\sigma})$ 

&\chm  &\chm &\chm & 
$Q_{G^2B^2}^{(3)} $ & $ (B_{\mu\nu} G^{A\mu\nu}) (B_{\rho\sigma} G^{A\rho\sigma})$
&\chm  &\chm &\chm 
\\
$Q_{W^4}^{(2)} $ & $ (W_{\mu\nu}^I \widetilde{W}^{I\mu\nu}) (W_{\rho\sigma}^J \widetilde{W}^{J\rho\sigma})$
&\chm  &\chm &\chm & 
$Q_{G^2B^2}^{(4)} $ & $ (B_{\mu\nu} \widetilde{G}^{A\mu\nu}) (B_{\rho\sigma} \widetilde{G}^{A\rho\sigma})$
&\chm  &\chm &\chm 
\\
$Q_{W^4}^{(3)} $ & $ (W_{\mu\nu}^I W^{J\mu\nu}) (W_{\rho\sigma}^I W^{J\rho\sigma})$ 
&\chm  &\chm &\chm & 
  \cyan $Q_{G^2B^2}^{(5)} $ & \cyan $ (B_{\mu\nu} \widetilde{B}^{\mu\nu}) (G_{\rho\sigma}^A G^{A\rho\sigma})$
&\chm & \xm &\xm 
  \\
$Q_{W^4}^{(4)} $ & $ (W_{\mu\nu}^I \widetilde{W}^{J\mu\nu}) (W_{\rho\sigma}^I \widetilde{W}^{J\rho\sigma})$ 
&\chm  &\chm &\chm & 
\cyan  $Q_{G^2B^2}^{(6)} $ &\cyan  $ (B_{\mu\nu} B^{\mu\nu}) (G_{\rho\sigma}^A \widetilde{G}^{A\rho\sigma})$
& \chm & \xm & \xm
  \\
 \cyan  $Q_{W^4}^{(5)} $ &\cyan  $ (W_{\mu\nu}^I W^{I\mu\nu}) (W_{\rho\sigma}^J \widetilde{W}^{J\rho\sigma})$ 
  &\chm & \xm & \xm & 
  \cyan $Q_{G^2B^2}^{(7)} $ & \cyan $ (B_{\mu\nu} G^{A\mu\nu}) (B_{\rho\sigma} \widetilde{G}^{A\rho\sigma})$
  &\chm &\xm &\xm
  \\
\cyan$Q_{W^4}^{(6)} $ & \cyan $ (W_{\mu\nu}^I W^{J\mu\nu}) (W_{\rho\sigma}^I \widetilde{W}^{J\rho\sigma})$ 
  & \chm & \xm &\xm & 
  $Q_{W^2B^2}^{(1)} $ & $ (B_{\mu\nu} B^{\mu\nu}) (W_{\rho\sigma}^I W^{I\rho\sigma})$
 &\chm  &\chm &\chm 
  \\
$Q_{B^4}^{(1)} $ & $ (B_{\mu\nu} B^{\mu\nu}) (B_{\rho\sigma} B^{\rho\sigma})$
 &\chm  &\chm &\chm & 
  $Q_{W^2B^2}^{(2)} $ & $ (B_{\mu\nu} \widetilde{B}^{\mu\nu}) (W_{\rho\sigma}^I \widetilde{W}^{I\rho\sigma})$
 &\chm  &\chm &\chm 
 \\
$Q_{B^4}^{(2)} $ & $ (B_{\mu\nu} \widetilde{B}^{\mu\nu}) (B_{\rho\sigma} \widetilde{B}^{\rho\sigma})$ 
 &\chm  &\chm &\chm & 
  $Q_{W^2B^2}^{(3)} $ & $ (B_{\mu\nu} W^{I\mu\nu}) (B_{\rho\sigma} W^{I\rho\sigma})$
 &\chm  &\chm &\chm 
\\
\cyan $Q_{B^4}^{(3)} $ &\cyan  $ (B_{\mu\nu} B^{\mu\nu}) (B_{\rho\sigma} \widetilde{B}^{\rho\sigma})$
& \chm & \xm & \xm & 
$Q_{W^2B^2}^{(4)} $ & $ (B_{\mu\nu} \widetilde{W}^{I\mu\nu}) (B_{\rho\sigma} \widetilde{W}^{I\rho\sigma})$
&\chm  &\chm &\chm 
\\  
$Q_{G^3B}^{(1)} $ & $ d^{ABC} (B_{\mu\nu} G^{A\mu\nu}) (G_{\rho\sigma}^B G^{C\rho\sigma})$
&\chm  &\chm &\chm & 
  \cyan  $Q_{W^2B^2}^{(5)} $ &\cyan  $ (B_{\mu\nu} \widetilde{B}^{\mu\nu}) (W_{\rho\sigma}^I W^{I\rho\sigma})$
  &\chm & \xm & \xm
\\
$Q_{G^3B}^{(2)} $ & $ d^{ABC} (B_{\mu\nu} \widetilde{G}^{A\mu\nu}) (G_{\rho\sigma}^B \widetilde{G}^{C\rho\sigma})$  
&\chm  &\chm &\chm & 
\cyan $Q_{W^2B^2}^{(6)} $ &\cyan  $ (B_{\mu\nu} B^{\mu\nu}) (W_{\rho\sigma}^I \widetilde{W}^{I\rho\sigma})$
&\chm & \xm & \xm
  \\
\cyan $Q_{G^3B}^{(3)} $ & \cyan $ d^{ABC} (B_{\mu\nu} \widetilde{G}^{A\mu\nu}) (G_{\rho\sigma}^B G^{C\rho\sigma})$  
&\chm & \xm & \xm & 
\cyan  $Q_{W^2B^2}^{(7)} $ &\cyan $ (B_{\mu\nu} W^{I\mu\nu}) (B_{\rho\sigma} \widetilde{W}^{I\rho\sigma})$
&\chm  &\xm  &\xm
  \\
\cyan $Q_{G^3B}^{(4)} $ &\cyan  $ d^{ABC} (B_{\mu\nu} G^{A\mu\nu}) (G_{\rho\sigma}^B \widetilde{G}^{C\rho\sigma})$
&\chm &\xm  &\xm &
& & & &
\\ \hline
\multicolumn{2}{|c|}{\boldmath$2:H^8$} & \multicolumn{3}{c|}{ }&  
\multicolumn{2}{c|}{\boldmath$4:H^4D^4$} & \multicolumn{3}{c|}{ }   \\ \hline
$Q_{H^8} $ & $ (H^\dag H)^4$  &\chm  &\chm &\chm &  
$Q_{H^4}^{(1)}$  &  $(D_{\mu} H^{\dag} D_{\nu} H) (D^{\nu} H^{\dag} D^{\mu} H)$
&\chm  &\chm &\chm  \\\cline{1-5}
\multicolumn{2}{|c|}{\boldmath$3:H^6D^2$} & \multicolumn{3}{c|}{ }  &
$Q_{H^4}^{(2)}$  &  $(D_{\mu} H^{\dag} D_{\nu} H) (D^{\mu} H^{\dag} D^{\nu} H)$  &\chm  &\chm &\chm  \\\cline{1-5}
$Q_{H^6}^{(1)}$  & $(H^{\dag} H)^2 (D_{\mu} H^{\dag} D^{\mu} H)$ &\chm  &\chm &\chm & 
$Q_{H^4}^{(3)}$  &  $(D^{\mu} H^{\dag} D_{\mu} H) (D^{\nu} H^{\dag} D_{\nu} H)$  &\chm  &\chm &\chm   \\
$Q_{H^6}^{(2)}$  & $(H^{\dag} H) (H^{\dag} \tau^I H) (D_{\mu} H^{\dag} \tau^I D^{\mu} H)$ &\chm  &\chm &\chm &   & & & & \\ \hline
\multicolumn{10}{|c|}{\boldmath$5:X^3H^2$}\\\hline
$Q_{G^3H^2}^{(1)} $ & $ f^{ABC} (H^\dag H) G_{\mu}^{A\nu} G_{\nu}^{B\rho} G_{\rho}^{C\mu}$    
&\chm  &\chm &\chm & 
\cyan  $Q_{W^3H^2}^{(2)} $ & \cyan $ \epsilon^{IJK} (H^\dag H) W_{\mu}^{I\nu}W_{\nu}^{J\rho} \widetilde{W}_{\rho}^{K\mu}$
&\chm &\xm & \xm \\
\cyan  $Q_{G^3H^2}^{(2)} $ &\cyan  $ f^{ABC} (H^\dag H) G_{\mu}^{A\nu} G_{\nu}^{B\rho} \widetilde{G}_{\rho}^{C\mu}$
&\chm & \xm &\xm &
$Q_{W^2BH^2}^{(1)}$ & $ \epsilon^{IJK} (H^\dag \tau^I H) B_{\mu}^{\,\nu} W_{\nu}^{J\rho} W_{\rho}^{K\mu}$
&\chm  &\chm &\chm  \\
$Q_{W^3H^2}^{(1)} $ & $\epsilon^{IJK} (H^\dag H) W_{\mu}^{I\nu}W_{\nu}^{J\rho} W_{\rho}^{K\mu}$
&\chm  &\chm &\chm & 
\cyan $Q_{W^2BH^2}^{(2)} $ &\cyan  $ \epsilon^{IJK} (H^\dag \tau^I H)(
\widetilde{B}^{\mu\nu} W_{\nu\rho}^J W_{\mu}^{K\rho}+{B}^{\mu\nu} W_{\nu\rho}^J \widetilde{W}_{\mu}^{K\rho}$
&\chm  &\xm &\xm  
\\\hline
\multicolumn{10}{c|}{\boldmath $6:X^2H^4$}  \\\hline
$Q_{G^2H^4}^{(1)} $ & $ (H^\dag H)^2 G^A_{\mu\nu} G^{A\mu\nu}$
&\chm  &\chm &\chm & 
\cyan $Q_{W^2H^4}^{(4)}$
& \cyan $(H^\dag \tau^I H) (H^\dag \tau^J H) \widetilde W^I_{\mu\nu} W^{J\mu\nu}$
&\chm &\xm &\xm \\
\cyan    $Q_{G^2H^4}^{(2)} $ &\cyan  $ (H^\dag H)^2 \widetilde G^A_{\mu\nu}G^{A\mu\nu}$
& \chm&\xm &\xm & 
$Q_{B^2H^4}^{(1)} $ & $  (H^\dag H)^2 B_{\mu\nu} B^{\mu\nu}$
&\chm  &\chm &\chm  \\
$Q_{W^2H^4}^{(1)} $ & $ (H^\dag H)^2 W^I_{\mu\nu} W^{I\mu\nu}$  
&\chm  &\chm &\chm & 
\cyan $Q_{B^2H^4}^{(2)}$ &  \cyan $(H^\dag H)^2 \widetilde B_{\mu\nu} B^{\mu\nu}$
&\chm &\xm &\xm \\
\cyan $Q_{W^2H^4}^{(2)}$ &  \cyan $(H^\dag H)^2 \widetilde W^I_{\mu\nu}W^{I\mu\nu}$   
& \chm &\xm &\xm &
$Q_{W^2H^4}^{(3)}$ & $(H^\dag \tau^I H) (H^\dag \tau^J H)W^I_{\mu\nu} W^{J\mu\nu}$  
&\chm  &\chm &\chm    \\
$Q_{WBH^4}^{(1)}$ &  $(H^\dag H) (H^\dag \tau^I H) W^I_{\mu\nu} B^{\mu\nu}$
&\chm  &\chm &\chm & 
\cyan  $Q_{WBH^4}^{(2)} $ & \cyan $ (H^\dag H) (H^\dag \tau^I H) \widetilde W^I_{\mu\nu} B^{\mu\nu}$
&\chm &\xm &\xm \\\hline
\multicolumn{10}{|c|}{\boldmath$7:X^2H^2D^2$}
\\\hline  
$Q_{G^2H^2D^2}^{(1)}$  &  $(D^{\mu} H^{\dag} D^{\nu} H) G_{\mu\rho}^A G_{\nu}^{A \rho}$ 
&\chm  &\chm &\chm & 
$Q_{B^2H^2D^2}^{(1)}$  &  $(D^{\mu} H^{\dag} D^{\nu} H) B_{\mu\rho} B_{\nu}^{\,\,\,\rho}$
&\chm  &\chm &\chm 
\\
$Q_{G^2H^2D^2}^{(2)}$  &  $(D^{\mu} H^{\dag} D_{\mu} H) G_{\nu\rho}^A G^{A \nu\rho}$ 
&\chm  &\chm &\chm & 
$Q_{B^2H^2D^2}^{(2)}$  &  $(D^{\mu} H^{\dag} D_{\mu} H) B_{\nu\rho} B^{\nu\rho}$ 
&\chm  &\chm &\chm  
\\
\cyan $Q_{G^2H^2D^2}^{(3)}$  & \cyan   $(D^{\mu} H^{\dag} D_{\mu} H) G_{\nu\rho}^A \widetilde{G}^{A \nu\rho}$
&\chm &\xm &\xm & 
\cyan $Q_{B^2H^2D^2}^{(3)}$  &  \cyan $(D^{\mu} H^{\dag} D_{\mu} H) B_{\nu\rho} \widetilde{B}^{\nu\rho}$
&\chm  &\xm &\xm  
\\
$Q_{W^2H^2D^2}^{(1)}$  &  $(D^{\mu} H^{\dag} D^{\nu} H) W_{\mu\rho}^I W_{\nu}^{I \rho}$
&\chm  &\chm &\chm & 
$Q_{WBH^2D^2}^{(1)}$  &  $(D^{\mu} H^{\dag} \tau^I D_{\mu} H) B_{\nu\rho} W^{I \nu\rho}$
&\chm  &\chm &\chm 
\\
$Q_{W^2H^2D^2}^{(2)}$  &  $(D^{\mu} H^{\dag} D_{\mu} H) W_{\nu\rho}^I W^{I \nu\rho}$
&\chm  &\chm &\chm & 
\cyan $Q_{WBH^2D^2}^{(2)}$  &  \cyan $(D^{\mu} H^{\dag} \tau^I D_{\mu} H) B_{\nu\rho} \widetilde{W}^{I \nu\rho}$
&\chm &\xm &\xm
\\
\cyan $Q_{W^2H^2D^2}^{(3)}$  & \cyan  $(D^{\mu} H^{\dag} D_{\mu} H) W_{\nu\rho}^I \widetilde{W}^{I \nu\rho}$
& \chm & \xm &\xm &
\cyan $Q_{WBH^2D^2}^{(3)}$  &  \cyan $i (D^{\mu} H^{\dag} \tau^I D^{\nu} H) (B_{\mu\rho} W_{\nu}^{I \rho}- B_{\nu\rho} W_{\mu}^{I\rho})$
&\xm &\chm & \xm
\\
$Q_{W^2H^2D^2}^{(4)}$  &  $i \epsilon^{IJK} (D^{\mu} H^{\dag} \tau^I D^{\nu} H) W_{\mu\rho}^J W_{\nu}^{K \rho}$
&\chm  &\chm &\chm & 
$Q_{WBH^2D^2}^{(4)}$  &  $(D^{\mu} H^{\dag} \tau^I D^{\nu} H) (B_{\mu\rho} W_{\nu}^{I \rho}+ B_{\nu\rho} W_{\mu}^{I\rho})$
&\chm  &\chm &\chm 
\\
\cyan $Q_{W^2H^2D^2}^{(5)}$  &\cyan   $\epsilon^{IJK} (D^{\mu} H^{\dag} \tau^I D^{\nu} H) (W_{\mu\rho}^J \widetilde{W}_{\nu}^{K \rho}- \widetilde{W}_{\mu\rho}^J W_{\nu}^{K \rho})$
&\chm &\xm  &\xm & 
\cyan $Q_{WBH^2D^2}^{(5)}$  &  \cyan $i (D^{\mu} H^{\dag} \tau^I D^{\nu} H) (B_{\mu\rho} \widetilde{W}_\nu^{^I \rho}- B_{\nu\rho} \widetilde{W}_\mu^{^I \rho})$
&\xm & \xm & \chm
\\
\cyan $Q_{W^2H^2D^2}^{(6)}$  &\cyan   $i \epsilon^{IJK} (D^{\mu} H^{\dag} \tau^I D^{\nu} H) (W_{\mu\rho}^J \widetilde{W}_{\nu}^{K \rho}+ \widetilde{W}_{\mu\rho}^J W_{\nu}^{K \rho})$
& \chm &\xm & \xm& 
\cyan $Q_{WBH^2D^2}^{(6)}$  &  \cyan $ (D^{\mu} H^{\dag} \tau^I D^{\nu} H) (B_{\mu\rho} \widetilde{W}_\nu^{^I \rho}+ B_{\nu\rho} \widetilde{W}_\mu^{^I \rho})$
&\chm &\xm & \xm \\ \hline
\multicolumn{10}{|c|}{\boldmath$8:XH^4D^2$}\\\hline
$Q_{WH^4D^2}^{(1)}$  & $(H^{\dag} H) (D^{\mu} H^{\dag} \tau^I D^{\nu} H) W_{\mu\nu}^I$
&\chm  &\chm &\chm & 
\cyan $Q_{WH^4D^2}^{(4)}$  & \cyan $\epsilon^{IJK} (H^{\dag} \tau^I H) (D^{\mu} H^{\dag} \tau^J D^{\nu} H) \widetilde{W}_{\mu\nu}^K$
&\xm &\xm & \chm\\
\cyan $Q_{WH^4D^2}^{(2)}$  & \cyan $(H^{\dag} H) (D^{\mu} H^{\dag} \tau^I D^{\nu} H) \widetilde{W}_{\mu\nu}^I$
&\chm &\xm &\xm &
$Q_{BH^4D^2}^{(1)}$  & $(H^{\dag} H) (D^{\mu} H^{\dag} D^{\nu} H) B_{\mu\nu}$
&\chm & \chm &\chm \\
\cyan $Q_{WH^4D^2}^{(3)}$  & \cyan $\epsilon^{IJK} (H^{\dag} \tau^I H) (D^{\mu} H^{\dag} \tau^J D^{\nu} H) W_{\mu\nu}^K)$
&\xm & \chm & \xm & 
\cyan $Q_{BH^4D^2}^{(2)}$ \cyan  & \cyan $(H^{\dag} H) (D^{\mu} H^{\dag} D^{\nu} H) \widetilde{B}_{\mu\nu}$
& \chm & \xm & \xm \\\hline
\end{tabular}}
\caption{89 purely bosonic operators present in Murphy's basis.}
\label{tab:89murphy}
\end{table}
%%%%%%%%%%%%%%%%%%%%%%%%%%%%%%%%%%%%%%%%%%%%%%%%%%%%%%%
\begin{table}
\scalebox{0.85}{
\begin{tabular}[t]{|l|l|c|c|c||l|l|c|c|c|}
\hline    
\multicolumn{2}{|c|}{\boldmath  $H^6 D^2$} & C & P & CP &
\multicolumn{2}{|c|}{\boldmath  $H^6 D^2$} & C & P & CP 
              \\
\hline
$R^{\prime(1)}_{H^6D^2}$ &   $(D^2 H^\dagger H + H^\dagger D^2 H)( H^\dagger H)( H^\dagger H)$
&\chm & \chm &\chm & 
\cyan $R^{\prime (2)}_{H^6D^2}$ & \cyan $i (H^\dagger D^2H-D^2 H^\dagger H) (H^\dagger H)( H^\dagger H)$
&\xm & \chm &\xm  
\\
\hline
\multicolumn{10}{|c|}{\boldmath$H^4 D^4$}\\\hline
$R^{\prime(1)}_{H^4D^4}$ &     $(D^2 H^\dagger \tau^I  H+ H^\dagger \tau^I D^2 H ) (D^\mu H^\dagger \tau^I D_\mu  H)$
&\chm & \chm &\chm & 
\cyan $R^{\prime(6)}_{H^4D^4}$ &  \cyan   i $(H^\dagger D^2 H -D^2 H^\dagger H)  (D^\mu H^\dagger D_\mu H)$
&\xm  & \chm & \xm 
\\
$R^{\prime (2)}_{H^4D^4}$ &     $(D^2 H^\dagger D_\mu  H)  ( H^\dagger D^\mu  H)+(D_\mu H^\dagger D^2  H) (D^\mu H^\dagger  H)$
&\chm & \chm &\chm & 
$R^{(7)}_{H^4D^4}$ &       $(D^2 H^\dagger D^2  H)( H^\dagger  H)$
&\chm & \chm &\chm 
\\
\cyan $R^{\prime(3)}_{H^4D^4}$ & \cyan     $i \left((D_\mu H^\dagger D^2  H) (D^\mu H^\dagger  H)-(D^2 H^\dagger D_\mu  H)  ( H^\dagger D^\mu  H)\right) $
& \xm &\chm &\xm & 
$R^{\prime (8)}_{H^4D^4}$ &     $(D^2 H^\dagger  H)(D^2 H^\dagger H)+( H^\dagger D^2  H)  ( H^\dagger D^2  H)$
&\chm & \chm &\chm 
\\
\cyan $R^{\prime (4)}_{H^4D^4}$ &  \cyan     $i(H^\dagger \tau^I D^2  H-D^2 H^\dagger \tau^I  H)  (D_\mu H^\dagger \tau^I  D^\mu  H)$
&\xm & \chm &\xm & 
$R^{(9)}_{H^4D^4}$ &      $(D^2 H^\dagger  H)  ( H^\dagger D^2  H)$
&\chm & \chm &\chm 
\\
$R^{\prime (5)}_{H^4D^4}$ &      $(D^2 H^\dagger   H+ H^\dagger D^2 H) (D_\mu  H^\dagger D^\mu  H )$
&\chm & \chm &\chm & 
\cyan $R^{\prime(10)}_{H^4D^4}$ & \cyan     $i\left(( H^\dagger D^2  H)  ( H^\dagger D^2  H)-(D^2 H^\dagger  H)(D^2 H^\dagger H)\right)$
&\xm & \chm &\xm 
\\ \hline
\multicolumn{2}{|c|}{\boldmath$H^2 D^6$} & \multicolumn{3}{|c||}{ }
& \multicolumn{2}{c|}{\boldmath$X^2 D^4$} & \multicolumn{3}{|c|}{ }
\\\hline
$R^{(1)}_{H^2D^6}$ &  $(D^\mu D^2  H^\dagger D^\mu D^2  H)$
&\chm & \chm &\chm  & 
$R^{(1)}_{B^2D^4}$ &  $D^\rho D^{\alpha}B_{\alpha\mu} D_\rho D^{\beta}B_{\beta}^{\mu}$
&\chm & \chm &\chm
\\
&
& & & & 
$R^{(1)}_{W^2D^4}$ & $D^\rho D^{\alpha}W^I_{\alpha\mu} D_\rho D^{\beta}W_{\beta}^{I,\mu}$ 
&\chm & \chm &\chm  \\
&
& & & &  
$R^{(1)}_{G^2D^4}$ & $D^\alpha D^\mu G^A_{\mu\nu} D_\alpha D^\rho G^{A,\;\nu}_{\rho}$   
&\chm & \chm &\chm  
\\\hline
\multicolumn{10}{|c|}{\boldmath  $X^3 D^2$, $X^2X' D^2$} 
\\
\hline
$R^{(1)}_{W^3D^2}$ & ${W^I_{\mu\nu} (D_{\alpha}W^{J,\alpha \mu})(D_{\beta}W^{K,\beta \nu})\epsilon^{IJK}}$
&\chm & \chm &\chm & 
$R^{(1)}_{G^3D^2}$ &  $G^A_{\mu\nu} (D_{\alpha}G^{B,\alpha\mu})(D_{\beta}G^{C,\beta \nu}) f^{ABC}$
&\chm & \chm &\chm
\\
\cyan $R^{(2)}_{W^3D^2}$ &\cyan ${\widetilde{W}^I_{\mu\nu} (D_{\alpha}W^{J,\alpha \mu})(D_{\beta}W^{K,\beta \nu})\epsilon^{IJK}}$
&\chm & \xm & \xm & 
\cyan $R^{(2)}_{G^3D^2}$ &  \cyan ${\widetilde{G}^A_{\mu\nu} (D_{\alpha}G^{B,\alpha\mu})(D_{\beta}G^{C,\beta \nu}) f^{ABC}}$
&\chm &\xm  &\xm 
\\
$R^{(3)}_{W^3D^2}$ & ${W^I_{\mu \nu} W^{J,\nu}_{\rho} (D^{\mu}D_{\alpha}W^{K,\alpha \rho}) \epsilon^{IJK}}$
&\chm & \chm &\chm & 
$R^{(3)}_{G^3D^2}$ & ${G^A_{\mu \nu} G^{B,\nu}_{\rho} (D^{\mu}D_{\alpha}G^{C,\alpha \rho}) f^{ABC}}$
&\chm & \chm &\chm
\\
\cyan $R^{(4)}_{W^3D^2}$ &  \cyan ${W^I_{\mu \nu} \widetilde{W}^{J,\nu}_{\rho} (D^{\mu}D_{\alpha}W^{K,\alpha \rho}- D^{\rho}D_{\alpha}W^{K,\alpha \mu}) \epsilon^{IJK}}$
&\chm &\xm  &\xm  & 
\cyan $R^{(4)}_{G^3D^2}$ &\cyan  ${G^A_{\mu \nu} \widetilde{G}^{B,\nu}_{\rho} (D^{\mu}D_{\alpha}G^{C,\alpha \rho}- D^{\rho}D_{\alpha}G^{C,\alpha \mu}) f^{ABC}}$
&\chm &\xm  &\xm 
\\
\cyan $R^{(1)}_{BW^2D^2}$ &\cyan  $(D^\mu B_{\mu\nu}) W^{I,\nu\rho} (D^\alpha W^{I}_{\rho\alpha})$
& \xm & \chm &\xm & 
\cyan $R^{(1)}_{BG^2D^2}$ &  \cyan  ${G^A_{\mu\nu} (D_{\alpha}G^{A,\alpha\mu})(D_{\beta}B^{\beta \nu})}$
&\xm  &\chm & \xm 
\\  
\cyan $R^{(2)}_{BW^2D^2}$ & \cyan  $(D^\mu B_{\mu\nu}) \widetilde{W}^{I,\nu\rho} (D^\alpha W^{I}_{\rho\alpha})$
&\xm  &\xm & \chm &  
\cyan  $R^{(2)}_{BG^2D^2}$ &\cyan  ${\widetilde{G}^A_{\mu\nu} (D_{\alpha}G^{A,\alpha\mu})(D_{\beta}B^{\beta \nu})}$
&\xm  &\xm & \chm 
\\   
\cyan $R^{(3)}_{BW^2D^2}$ & \cyan ${B_{\mu \nu} W^{I,\nu}_{\rho} (D^{\mu}D_{\alpha}W^{I,\alpha \rho} - D^{\rho}D_{\alpha}W^{I,\alpha \mu})}$
&\xm  &\chm & \xm & 
\cyan $R^{(3)}_{BG^2D^2}$ & \cyan ${B_{\mu \nu} G^{A,\nu}_{\rho} (D^{\mu}D_{\alpha}G^{A,\alpha \rho} - D^{\rho}D_{\alpha}G^{A,\alpha \mu})}$
&\xm  &\chm & \xm 
\\
\cyan $R^{(4)}_{BW^2D^2}$ &\cyan${B_{\mu \nu} \widetilde{W}^{I,\nu}_{\rho}(D^{\mu}D_{\alpha}W^{I,\alpha \rho} - D^{\rho}D_{\alpha}W^{I,\alpha \mu})}$
&\xm  &\xm & \chm  & 
\cyan $R^{(4)}_{BG^2D^2}$ &  \cyan ${B_{\mu \nu} \widetilde{G}^{A,\nu}_{\rho} (D^{\mu}D_{\alpha}G^{A,\alpha \rho} - D^{\rho}D_{\alpha}G^{A,\alpha \mu})}$
&\xm  &\xm & \chm 
\\\hline
\multicolumn{10}{|c|}{\boldmath  $X^2H^2 D^2$}
\\\hline
$R^{\prime (1)}_{B^2H^2 D^2}$ & $B_{\mu\nu} B^{\mu\nu} (D^2H^\dagger H + H^\dagger D^2 H)$
&\chm & \chm &\chm & 
$R^{\prime (1)}_{G^2H^2 D^2}$ & $G^{A}_{\mu\nu} G^{A\mu\nu} (D^2H^\dagger H+ H^\dagger D^2 H)$
&\chm & \chm &\chm
\\
\cyan $R^{\prime (2)}_{B^2H^2 D^2}$ & \cyan $ i B_{\mu\nu} B^{\mu\nu} (H^\dagger D^2H-D^2 H^\dagger H)$
&\xm & \chm &\xm & 
\cyan $R^{(2)}_{G^2H^2 D^2}$ &
\cyan $i G^{A}_{\mu\nu} G^{A\mu\nu} (H^\dagger D^2H-D^2 H^\dagger H)$
&\xm & \chm &\xm  
\\
\cyan $R^{\prime (3)}_{B^2H^2 D^2}$ &\cyan $B_{\mu\nu} \widetilde{B}^{\mu\nu} (D^2H^\dagger H+H^\dagger D^2 H)$
& \chm & \xm & \xm & 
\cyan $R^{\prime (3)}_{G^2H^2 D^2}$ & \cyan $G^{A}_{\mu\nu} \widetilde{G}^{A\mu\nu} (D^2H^\dagger H +H^\dagger D^2 H)$
& \chm & \xm & \xm 
\\
\cyan $R^{\prime (4)}_{B^2H^2 D^2}$ & \cyan $i B_{\mu\nu} \widetilde{B}^{\mu\nu} (H^\dagger D^2H-D^2H^\dagger H)$ 
& \xm &\xm  &\chm & 
\cyan $R^{\prime (4)}_{G^2H^2 D^2}$ & \cyan $i G^{A}_{\mu\nu} \widetilde{G}^{A\mu\nu} (H^\dagger D^2H-D^2H^\dagger H)$
& \xm &\xm  &\chm 
\\
$R^{\prime (5)}_{B^2H^2 D^2}$ & $(D^\mu B_{\mu\nu}) B^{\alpha\nu}  D_\alpha (H^\dagger H)$
&\chm & \chm &\chm & 
$R^{\prime (5)}_{G^2H^2 D^2}$ & $(D^\mu G^{A}_{\mu\nu}) G^{A\alpha\nu}  D_\alpha (H^\dagger  H )$
&\chm & \chm &\chm 
\\
\cyan $R^{\prime (6)}_{B^2H^2 D^2}$ & \cyan $i (D^\mu B_{\mu\nu}) B^{\alpha\nu}  (H^\dagger \overleftrightarrow{D}_\alpha H)$ 
&\xm & \chm& \xm& 
\cyan $R^{\prime (6)}_{G^2H^2 D^2}$ & \cyan $ i (D^\mu G^{A}_{\mu\nu}) G^{A\alpha\nu}  (H^\dagger \overleftrightarrow{D}_\alpha H)$ 
&\xm & \chm &\xm
\\
\cyan $R^{\prime (7)}_{B^2H^2 D^2}$ & \cyan $(D^\mu B_{\mu\nu}) \widetilde{B}^{\alpha\nu}  D_\alpha (H^\dagger H)$
&\chm &\xm &\xm & 
\cyan $R^{\prime (7)}_{G^2H^2 D^2}$ &\cyan $(D^\mu G^{A}_{\mu\nu}) \widetilde{G}^{A\alpha\nu}  D_\alpha (H^\dagger H)$
&\chm &\xm &\xm 
\\
\cyan $R^{\prime (8)}_{B^2H^2 D^2}$ & \cyan $i (D^\mu B_{\mu\nu}) \widetilde{B}^{\alpha\nu}  (H^\dagger \overleftrightarrow{D}_\alpha H)$ 
&\xm & \xm & \chm & 
\cyan $R^{\prime (8)}_{G^2H^2 D^2}$ &\cyan $i (D^\mu G^{A}_{\mu\nu}) \widetilde{G}^{A\alpha\nu}  (H^\dagger \overleftrightarrow{D}_\alpha H)$ 
&\xm & \xm & \chm 
\\
$R^{(9)}_{B^2H^2 D^2}$ & $(D^\mu B_{\mu\alpha})(D_\nu  B^{\nu\alpha})  (H^\dagger H)$
&\chm & \chm &\chm & 
$R^{(9)}_{G^2H^2 D^2}$ &     $(D^\mu G^{A}_{\mu\alpha})(D_\nu  G^{A\nu\alpha})  (H^\dagger H)$
&\chm & \chm &\chm
\\
$R^{\prime (1)}_{W^2H^2 D^2}$ & $W^I_{\mu\nu} W^{I,\mu\nu} (D^2H^\dagger H +H^\dagger D^2H)$
&\chm & \chm &\chm & 
$R^{\prime (1)}_{BWH^2 D^2}$ &$B_{\mu\nu} W^{I,\mu\nu} (H^\dagger \tau^I D^2H +H^\dagger\tau^I D^2H)$
&\chm & \chm &\chm
\\
\cyan $R^{\prime (2)}_{W^2H^2 D^2}$ &\cyan $i W^I_{\mu\nu} W^{I,\mu\nu} (H^\dagger D^2H-D^2H^\dagger H)$
& \xm & \chm &\xm & 
\cyan $R^{\prime (2)}_{BWH^2 D^2}$ & \cyan $i B_{\mu\nu} W^{I,\mu\nu} (D^2 H^\dagger \tau^I H-H^\dagger \tau^I D^2H)$
& \xm & \chm &\xm 
\\
\cyan $R^{\prime (3)}_{W^2H^2 D^2}$ & \cyan $W^I_{\mu\nu} \widetilde{W}^{I,\mu\nu} (D^2H^\dagger H +H^\dagger D^2H)$
&\chm &\xm &\xm & 
\cyan $R^{\prime (3)}_{BWH^2 D^2}$ & \cyan $B_{\mu\nu} \widetilde{W}^{I,\mu\nu} (H^\dagger \tau^I D^2H + D^2 H^\dagger \tau^I H)$
&\chm &\xm &\xm 
\\
\cyan $R^{\prime (4)}_{W^2H^2 D^2}$ & \cyan $i W^I_{\mu\nu} \widetilde{W}^{I,\mu\nu} (H^\dagger D^2H-D^2H^\dagger H)$
&\xm &\xm &\chm &
\cyan $R^{\prime (4)}_{BWH^2 D^2}$ & \cyan $i B_{\mu\nu} \widetilde{W}^{I,\mu\nu} (D^2 H^\dagger \tau^I H -H^\dagger \tau^I D^2H)$
&\xm &\xm &\chm 
\\
 $R^{\prime (5)}_{W^2H^2 D^2}$ & $(D^\mu W^I_{\mu\nu}) W^{I,\alpha\nu}  D_\alpha (H^\dagger H)$
&\chm & \chm &\chm & 
$R^{\prime (5)}_{BWH^2 D^2}$ & $(D^\mu B_{\mu\alpha}) W^{I,\alpha\nu}  D_\nu (H^\dagger \tau^I H)$
&\chm & \chm &\chm
\\
\cyan $R^{\prime (6)}_{W^2H^2 D^2}$ & \cyan $i (D^\mu W^I_{\mu\nu}) W^{I,\alpha\nu}  (H^\dagger \overleftrightarrow{D}_\alpha H)$
& \xm & \chm & \xm & 
\cyan $R^{\prime (6)}_{BWH^2 D^2}$ & \cyan $i (D^\mu B_{\mu\alpha}) W^{I,\alpha\nu}  (H^\dagger \overleftrightarrow{D}^I_\nu H)$
& \xm & \chm & \xm 
\\
\cyan $R^{\prime (7)}_{W^2H^2 D^2}$ & \cyan $(D^\mu W^I_{\mu\nu}) \widetilde{W}^{I,\alpha\nu}  D_\alpha (H^\dagger H)$
& \chm & \xm & \xm &
\cyan $ R^{\prime (7)}_{BWH^2 D^2}$ & \cyan $ (D^\mu B_{\mu\alpha}) \widetilde{W}^{I,\alpha\nu}  D_\nu (H^\dagger \tau^I H)$
& \chm & \xm & \xm 
\\
\cyan $ R^{(\prime 8)}_{W^2H^2 D^2}$ & \cyan $i (D^\mu W^I_{\mu\nu}) \widetilde{W}^{I,\alpha\nu}  (H^\dagger \overleftrightarrow{D}_\alpha H)$
& \xm & \xm & \chm  & 
\cyan $R^{\prime (8)}_{BWH^2 D^2}$ &\cyan $i (D^\mu B_{\mu\alpha}) \widetilde{W}^{I,\alpha\nu}  (H^\dagger \overleftrightarrow{D}^I_\nu H)$
& \xm & \xm & \chm  
\\
$R^{(9)}_{W^2H^2 D^2}$ & $(D^\mu W^I_{\mu\alpha}) (D_\nu W^{I,\nu\alpha})  (H^\dagger H) $
&\chm & \chm &\chm & 
$R^{\prime (9)}_{BWH^2 D^2}$ & $(D^\mu  W^I_{\mu\nu}) B^{\nu\alpha} D_\alpha (H^\dagger\tau^I H)$
&\chm & \chm &\chm 
\\
\cyan $R^{\prime (10)}_{W^2H^2 D^2}$ & \cyan $\epsilon^{IJK} (D^\mu W^I_{\mu\nu}) W^{J,\rho\nu} D_\rho( H^\dagger \tau^K H)$
& \xm & \chm & \xm & 
\cyan $R^{\prime (10)}_{BWH^2 D^2}$ & \cyan $i (D^\mu  W^I_{\mu\nu}) B^{\nu\alpha} (H^\dagger \overleftrightarrow{D}^I_\alpha H)$
&\xm  &\chm  &\xm 
\\
 $R^{\prime (11)}_{W^2H^2 D^2}$ & $i \epsilon^{IJK} (D^\mu W^I_{\mu\nu}) W^{J,\rho\nu} (H^\dagger \overleftrightarrow{D}^K_\rho H)$
&\chm & \chm &\chm & 
\cyan $R^{\prime (11)}_{BWH^2 D^2}$ & \cyan $(D^\mu  W^I_{\mu\nu}) \widetilde{B}^{\nu\alpha} D_\alpha(H^\dagger \tau^I H)$
&\chm &\xm & \xm
\\
\cyan  $R^{\prime (12)}_{W^2H^2 D^2}$ & \cyan $ \epsilon^{IJK} (D^\mu W^I_{\mu\nu}) \widetilde{W}^{J,\rho\nu} D_\rho (H^\dagger \tau^K H)$
&\xm &\xm &\chm &  
\cyan  $R^{\prime (12)}_{BWH^2 D^2}$ & \cyan $i (D^\mu  W^I_{\mu\nu}) \widetilde{B}^{\nu\alpha} (H^\dagger \overleftrightarrow{D}^I_\alpha H)$
&\xm &\xm &\chm 
\\
\cyan $R^{\prime (13)}_{W^2H^2 D^2}$ & \cyan  $i \epsilon^{IJK} (D^\mu W^I_{\mu\nu}) \widetilde{W}^{J,\rho\nu} (H^\dagger \overleftrightarrow{D}^K_\rho H)$  
&\chm & \xm & \xm & 
$R^{(13)}_{BWH^2 D^2}$ & $(D^\mu B_{\mu\alpha})(D_\nu  W^{I,\nu\alpha})  (H^\dagger\tau^I H)$
& \chm &\chm &\chm  
\\\hline
\multicolumn{2}{|c|}{\boldmath$X H^2 D^4$} & \multicolumn{3}{|c||}{} &
\multicolumn{2}{c|}{\boldmath$X H^4 D^2$}  & \multicolumn{3}{|c|}{}
\\\hline
\cyan $R^{\prime(1)}_{BH^2D^4}$ &  \cyan ${ (D^\mu H^\dagger D^2H+D^2H^\dagger { D^\mu H}) (D^\nu B_{\mu\nu} ) }$
&\xm  &\chm  &\xm & 
$R^{(1)}_{BH^4D^2}$ &  $i (D_{\alpha}B^{\alpha \mu})(H^\dagger \overleftrightarrow{D}_{\mu} H) (H^\dagger H)$
&\chm  &\chm  &\chm  
\\
$R^{\prime (2)}_{BH^2D^4}$ & $i ({D^\mu H^\dagger D^2H - D^2H^\dagger { D^\mu H})(D^\nu B_{\mu\nu} )  }$
&\chm  &\chm  &\chm  &
$R^{(1)}_{WH^4D^2}$ & $i (D^\mu W^I_{\mu\nu}) (H^\dag \overleftrightarrow{D}^{I \nu} H)  (H^\dag H)$
&\chm  &\chm  &\chm  
\\
$R^{(3)}_{BH^2D^4}$  &  ${i (D_\alpha D^\nu B_{\mu\nu} ) (D^\mu H^\dagger D^\alpha H- D^\alpha H^\dagger D^\mu H}$
&\chm  &\chm  &\chm  &  
$R^{\prime (2)}_{WH^4D^2}$ & $\epsilon^{IJK}(H^\dag \tau^I H) D^{\nu}(H^\dag \tau^{J} H) (D^{\mu}W^K_{\mu\nu})$
&\chm  &\chm  &\chm  
\\
\cyan $R^{\prime (1)}_{WH^2D^4}$ & \cyan ${ (D^\mu H^\dagger \tau^I D^2H +D^2H^\dagger \tau^I D^\mu H) (D^\nu W^I_{\mu\nu} ) }$
&\xm & \chm & \xm & 
\cyan $R^{\prime (3)}_{WH^4D^2}$ & \cyan $i \epsilon^{IJK} (H^\dag \tau^I H) (H^\dag \overleftrightarrow{D}^{J\nu} H) (D^{\mu}W^K_{\mu\nu})$
&\xm  &\chm  &\xm
\\
$R^{\prime (2)}_{WH^2D^4}$ & ${  (D^\mu H^\dagger \tau^I D^2H +D^2H^\dagger \tau^I D^\mu H) (D^\nu W^I_{\mu\nu} )  }$
&\chm  &\chm  &\chm  &
& & & &  
\\
$R^{(3)}_{WH^2D^4}$ & $i {(D_\alpha D^\nu W^I_{\mu\nu} ) (D^\mu H^\dagger\tau^I D^\alpha H-D^\alpha H^\dagger\tau^I D^\mu H)}$
&\chm  &\chm  &\chm   &
& & & &  
\\\hline
\end{tabular}}\\
\caption{Additional 86 bosonic operators present in universal theories}
\label{tab:redundant}
\end{table}

\section{Oblique parameters from dimension-six and dimension-eight operators}
\label{app:oblique}
At linear order in the coefficients of dimension-six and
dimension-eight operators in SMEFT eight non-zero oblique parameters
are generated.
We first compute them explicitly in terms of the coefficients
of the operators in the bosonic basis. Subsequently we express the results
in terms of the coefficients in the rotated basis by applying the relations in
Eqs.~\eqref{eq:rot6} and ~\eqref{eq:rot8}.
\begin{eqnarray}
  \hat S&=&\frac{\ch}{\sh}\frac{\vh^2}{\Lambda^2} \left[b_{BW}
    +\frac{\vh^2}{\Lambda^2} \left(
    b_{WBH^4}^{(1)}-\frac{g'}{4}r^{(1)}_{WH^4D^2}-\frac{g}{4}r^{(1)}_{BH^4D^2}\right)
    \right] \nonumber\\
  &=&\left[\frac{\ch}{\sh} \overline{c}_{BW}-\frac{\eh^2}{2\sh^2}
    \Big(\overline{c}_{2JW}+\overline{c}_{2JB}\Big)\right]\frac{\vh^2}{\Lambda^2}
  -\left[\frac{\eh^2}{8\ch\sh^3}c^{(7)}_{\psi^4H^2}
    +\frac{\eh^4}{8\sh^4\ch^2}
    \Big(\sh^2 c^{(2)}_{\psi^4 D^2}+\ch^2c^{(3)}_{\psi^4 D^2}\Big)
    \right]\frac{\vh^4}{\Lambda^4}
    \nonumber\\[8pt]
\hat
T&=&-\frac{\vh^2}{2\Lambda^2}
\left(b_{\Phi,1}+\frac{\vh^2}{\Lambda^2}b_{H^6}^{(2)}\right)   =
-\frac{1}{2}\left[\overline{c}_{\Phi,1}+\frac{\eh^2}{\ch^2}\overline{c}_{2JB}\right]
\frac{\vh^2}{\Lambda^2}
-\frac{\eh^2}{2\sh_2}c^{(7)}_{\psi^4H^2}
\frac{\vh^4}{\Lambda^4}
\nonumber\\[8pt]
W&=&\frac{g^2\vh^2}{4\Lambda^2}\left[r_{2W}-\frac{\vh^2}{\Lambda^2}
  r_{W^2H^2D^2}^{(9)}\right]=
-\frac{\eh^2}{2\sh^2} \overline{c}_{2JW}
{ \frac{\vh^2}{\Lambda^2}}= \overline{\Delta}_{4F}{ \frac{\vh^2}{\Lambda^2}}
\nonumber\\[8pt]
Y&=&\frac{g^2\vh^2}{4\Lambda^2}\left[r_{2B}-\frac{\vh^2}{\Lambda^2}
  r_{B^2H^2D^2}^{(9)}\right]=
-\frac{\eh^2}{2\sh^2} \overline{c}_{2JB}
{ \frac{\vh^2}{\Lambda^2}}
\label{eq:obliquedim8}
\\[8pt]
\hat U&=&\frac{\vh^4}{\Lambda^4}\left[b_{W^2H^4}^{(3)}+\frac{g}{2}
  r^{'(2)}_{WH^4D^2}\right]=
{ \left[\overline{c}_{W^2H^4}^{(3)} - \frac{\eh^2}{2\sh_2} c^{(7)}_{\psi^4H^2}-\frac{\eh^4}{2\sh_2^2} c^{(2)}_{\psi^4D^2} \right]   \frac{\vh^4}{\Lambda^2}}
\nonumber\\[8pt]
X&=&\frac{g^2\vh^4}{8\Lambda^4} r_{BWH^2D^2}^{(13)}=
\frac{\eh^2}{8\sh^2} c^{(7)}_{\psi^4H^2}
{ \frac{\vh^4}{\Lambda^4}}
\nonumber\\[8pt]
W'&=&-\frac{g^4\vh^4}{8\Lambda^4} r_{W^2D^4}^{(1)}=
-\frac{\eh^4}{8\sh^4} c^{(3)}_{\psi^4D^2}
{ \frac{\vh^4}{\Lambda^4}}
\nonumber\\[8pt]
Y'&=&-\frac{g^4\vh^4}{8\Lambda^4} r_{B^2D^4}^{(1)}=
-\frac{\eh^4}{8\sh^4} c^{(2)}_{\psi^4D^2}
{ \frac{\vh^4}{\Lambda^4}}
\nonumber
\\[8pt]
X'&=&V=V'=0\nonumber
\end{eqnarray}

\end{document}